\documentclass[pra,twocolumn,superscriptaddress,10pt]{revtex4-1}
\usepackage{epsfig}
\usepackage{epstopdf}
\usepackage{dsfont}
\usepackage{graphicx}
\usepackage{dcolumn}
\usepackage{color}
\usepackage{times}
\usepackage{bm}
\usepackage{amssymb}
\usepackage{amsmath}
\usepackage{epsfig}
\usepackage{epstopdf}
\usepackage{dsfont}
\usepackage{subfigure}

\begin{document}
\renewcommand{\thefootnote}{\fnsymbol {footnote}}

\title{Quantum-memory-assisted entropic uncertainty relations}

\author{Dong Wang} \email{dwang@ahu.edu.cn}
\affiliation{School of Physics \& Material Science, Anhui University, Hefei
230601,  People's Republic of China}
\affiliation{CAS  Key  Laboratory  of
Quantum  Information,  University  of  Science  and
Technology of China, Hefei 230026,  People's Republic of China}

\author{Fei Ming}
\affiliation{School of Physics \& Material Science, Anhui University, Hefei
230601,  People's Republic of China}

\author{Ming-Liang Hu} \email{mlhu0301@163.com}
\affiliation{School of Science, Xi'an University of Posts and Telecommunications, Xi'an 710121, China}

\author{Liu Ye} \email{yeliu@ahu.edu.cn}
\affiliation{School of Physics \& Material Science, Anhui University, Hefei
230601,  People's Republic of China}

%\author[D. Wang]{Dong Wang \inst{1,2,}\footnote{Corresponding author \quad E-mail:~\textsf{dwang@ahu.edu.cn}}}
%\author[F.\, Ming]{Fei Ming \inst{1}}
%\author[M.-L.\, Hu]{Ming-Liang Hu\inst{3,}\footnote{Corresponding author \quad E-mail:~\textsf{mlhu0301@163.com}}}
%\author[L.\, Ye]{Liu Ye\inst{1,}\footnote{Corresponding author \quad E-mail:~\textsf{yeliu@ahu.edu.cn}}}
%
%\address[1]{School of Physics \& Material Science, Anhui University, Hefei 230601, China}
%\address[2]{CAS Key Laboratory of Quantum Information, University of Science and Technology of China, Hefei 230026, China}
%\address[3]{School of Science, Xi'an University of Posts and Telecommunications, Xi'an 710121, China}

%\shortauthors{D. Wang et al.}
%%% Shortened author names for the column titles. This is necessary only if
%%% there are more than two authors.

%%% In two-column mode, abstracts are typeset inside a colored box. If the
%%% abstract text fits nicely in one column it should be typeset that way. This
%%% is achieved by giving the \shortabstract directive. If your abstract won't
%%% fit (or if you are in doubt if it will once the final typefaces are applied)
%%% please leave that decision to the editor.

\begin{abstract}{\bf
Uncertainty relations take a crucial and fundamental part in the frame of
quantum theory, and are bringing on many marvelous applications in the
emerging field of quantum information sciences. Especially, as entropy is
imposed into the uncertainty principle, entropy-based uncertainty relations
lead to a number of applications including quantum key distribution,
entanglement witness, quantum steering, quantum metrology, and quantum
teleportation. Herein, the history of the development of the uncertainty
relations is discussed, especially focusing on the recent progress with regard
to quantum-memory-assisted entropic uncertainty relations and dynamical
characteristics of the measured uncertainty in some explicit physical systems.
The aims are to help deepen the understanding of entropic uncertainty
relations and prompt further explorations for versatile applications of the
relations on achieving practical quantum tasks.}
\end{abstract}

%\shortabstract
%%% Here, the document begins.

\maketitle

\section{Introduction} \label{sec:1}
In the field of quantum mechanics, the uncertainty principle was recognized as one of the most fundamental and important features, remarkably differing from its classical counterpart. Originally, Heisenberg \cite{Heisenberg} proposed the famous uncertainty relation in 1927, related to momentum and position measured for a particle, later been rigorously proven by Kennard \cite{E.H.Kennard}. Both the measurement outcomes of the quantities cannot be predicted simultaneously and precisely, {\it i.e.}, the certainty of the estimation for the position of a particle implies the uncertainty of the estimation for its momentum, and vice versa. Actually, apart from moment and position, such a limitation also applies to phases and excitation numbers of harmonic oscillators, orthogonal components of spin angular momentum, and angle and orbital angular momentum of a particle \cite{PatrickJ}.

When considering arbitrary two observables, Robertson formulated a general formula via variation from the Heisenberg uncertainty relation (see Sec. 2.1). The standard deviation substantively provides a nontrivial tradeoff between two incompatible measurements. However, it has a drawback, lying in that the lower bound of Robertson's inequality relies on the concrete state of the system, bringing on a trivial result when the system is prepared in the eigenstates of anyone of the two observables. With the rise of quantum information theory, the notion of entropy was considered to be useful for formulating the uncertainty relation. Everett \cite{Everett1} and Hirschman \cite{Hirschman} originally put forward an entropy-based uncertainty relation regarding position and momentum observables. Subsequently, the improvement on this relation was obtained for arbitrary two non-commuting observables in Refs. \cite{Beckner} and \cite{Birula}.
{{In Sec. 2, the different types of uncertainty relations (in terms of variance, entropy and majorization) will be discussed in detail when the measured system is isolated from others.}}

Note that while the previous literatures only investigated the entropic uncertainty relation (EUR) for a single-partite system, two basic questions can be raised: what is the new expression for EUR if the system to be probed is correlated with another subsystem (say, a quantum memory) in a nonclassical way, and are there any new physical implications for these cases? These will be answered in Sec. 3.
 {{To precisely estimate the outcome of measurement, various optimized uncertainty's bounds will be introduced. Then quantum-memory-assisted EURs are generalized to the case of multiple measurement setting.}}

In reality, any quantum system is unavoidably susceptible to its ambient surroundings, and this will induce the phenomenon of decoherence and dissipation. Due to this fact, it is significant to make clear how the environment influences the uncertainty of a measurement in realistic quantum information processing tasks. Moreover, how to control the amount of the uncertainty ought to be basically interesting in the regime of quantum precision measurement.
All these issues will be reviewed in Sec. 4.

Based on the EURs in the presence of quantum memory, the uncertainty had produced versatile applications, including entanglement witness, quantum teleportation, quantum cryptography, quantum speedup, creating steering inequality and quantum metrology, and so forth (see Sec. 5). In Particular, quantum key distribution was commercialized in nowadays markets and its security roots in the Heisenberg's uncertainty principle.

Actually, there already exist several reviews about the theme of EUR. In 2010, Wehner and Winter \cite{Birula2} reviewed the EUR with respect to multi-discrete variable from the viewpoint of information theory, and lately Bia{\l}ynicki-Birula and Rudnicki \cite{Beckner2} reviewed the continuous variable EURs from the physics perspective. Besides, Coles \emph{et al.} \cite{PatrickJ} mainly summarized the previous two, and added the recent development in their applications. {{Very recently, Hertz and Cerf \cite{hc1} reviewed in detail continuous-variable entropic uncertainty relations.}} Different from the previous reviews, we will mainly contribute to the improved EURs with a quantum memory and the dynamics of the measured uncertainty and its control via various approaches.
%The structure of this review has been diagrammatized in Fig. \ref{f1}.
We thus aim to offer the recent progresses related to EUR, which might be helpful to facilitate the spread of their performance in quantum information and new quantum technologies.
%%%%%%%%%%%%%%%%%%%%%%%%%%%%
%\begin{figure}
%\centering
%\includegraphics[width=8.5cm]{f1.eps}
%\caption{Plan of this review. (Top) Sec. \ref{sec:2}: Uncertainty relations without a quantum memory; the inset shows the scheme to experimentally demonstrate the uncertainty relations based on the sum of variances proposed by Maccone and Pati \cite{Kunkun}. (Right) Sec. \ref{sec:3}: Quantum-memory-assisted EUR; the inset is an illustration of the "uncertainty game" [28]. (Bottom) Sec. \ref{sec:4}: The dynamics of the EUR; the inset describes the scheme to observe the dynamics of uncertainty in certain specific environment \cite{D.Wang}. (Left) Sec. \ref{sec:5}: Applications of the EUR; the inset shows role of the EUR in witnessing entanglement \cite{Nphys1}.}
%\label{f1}
%\end{figure}

\section {Uncertainty relation in the absence of a quantum memory} \label{sec:2}

In this section, we will focus on various uncertainty relations via variance and entropy when the measured system is isolated from others.

\subsection{Uncertainty relations based on variance}

Seminally, Heisenberg in 1927 proposed the celebrated uncertainty relation, showing that one is unable to capture simultaneously the precise measurement's outcomes with certainty for the position and momentum of a particle \cite{Heisenberg}, which usually can be expressed by the inequality $\Delta p\Delta x\geq \hbar/2$. Certainly, this derivation is deemed as the principal character in the regime of quantum physics different from its classical counterpart. As to two arbitrary incompatible observables ${Q}$ and ${R}$, Kennard \cite{E.H. Kennard} and Robertson \cite{H. P. Robertson} derived a standard deviation
%%%%%%%%%%%%%%%%%%%%%%%%%%%%
\begin{align}
 \Delta{ {Q}}\cdot\Delta{{ R}}\geq \frac12 \left|\langle[{{Q}},{ R}]\rangle\right|,
 \label{Eq.2-1}
\end{align}
%%%%%%%%%%%%%%%%%%%%%%%%%%%%
where $\Delta {{R}}=\sqrt{{\langle{R}^2\rangle -\langle{ R}\rangle^2}}$ denotes the variance of $R$, with $\langle R\rangle$ being the expectation value of the observable $R$, and $[Q,R]={ QR-RQ}$ gives the commutator of the operators $Q$ and $R$ \cite{qiao}. Therewith, Schr\"{o}dinger \cite{Schr} strengthened the Kennard-Robertson's result by means of appending an anti-commutator term, leading to
%%%%%%%%%%%%%%%%%%%%%%%%%%%%
\begin{align}
\Delta{ {Q}}^2\cdot\Delta{{ R}}^2\geq \left|\frac12\langle[{{Q}},{ R}]\rangle\right|^2+\left|\frac12\langle\{{{Q}},{ R}\}\rangle-\langle Q\rangle\langle R\rangle\right|^2,
\label{Eq.2-2}
\end{align}
%%%%%%%%%%%%%%%%%%%%%%%%%%%%
{{Nevertheless, the lower bounds in Eqs. (\ref{Eq.2-1}) and (\ref{Eq.2-2}) are state dependent. If the system is prepared in one of the eigenstates of $Q$ or $R$,
one can easily   work out  $\left| {\left\langle {\left[ {Q,\left. R \right]} \right.} \right\rangle } \right| = 0$ and $ \left|\frac 12 {\left\langle {\left\{ {\left[ {Q,\left. R \right]} \right.} \right\}} \right\rangle  - \left\langle Q \right\rangle \left\langle R \right\rangle } \right| = 0$, which naturally leads to the fact that the lower bounds in Eqs. (\ref{Eq.2-1}) and (\ref{Eq.2-2}) will be zero-valued. This means that the standard deviations will become ineffective and trivial to measure the uncertainty in such a situation.}} Until recently, Maccone and Pati \cite{Lorenzo} removed this drawback and proposed a strong uncertainty relations, reading as
%%%%%%%%%%%%%%%%%%%%%%%%%%%%
\begin{align}
\Delta{ {Q}}^2+\Delta{{ R}}^2\geq  \max\left\{{\cal B}_1, {\cal B}_2\right\},
\label{Eq.2-3}
\end{align}
%%%%%%%%%%%%%%%%%%%%%%%%%%%%
with
%%%%%%%%%%%%%%%%%%%%%%%%%%%%
\begin{equation}
 \begin{split}
   & {\cal B}_1=\pm i\langle[{{Q}},{ R}]\rangle+\left|\langle\Psi|Q\pm iR|\Psi^\bot\rangle\right|^2 \\
   & {\cal B}_2=\frac12\left|\langle\Psi^\bot_{Q+R}\left|Q+R\right|\Psi\rangle\right|^2 \\
 \end{split}
\end{equation}
%%%%%%%%%%%%%%%%%%%%%%%%%%%%
and the state $|\Psi^\bot\rangle$ is orthogonal to $|\Psi\rangle$. The relation of Eq. (\ref{Eq.2-3}) has also been demonstrated by some promising experiments \cite{Kunkun,Kunkun2,Kunkun3}.

\subsection{Uncertainty relations based on entropy}

Technically, there is other working and straight approach to depict the uncertainty relations by the concept of entropy rather than the deviation mentioned above.

\subsubsection{Uncertainty relation based on differential entropy}

Beckner \cite{Beckner} as well as Bia{\l}ynicki-Birula and Mycielski \cite{Birula} reported the uncertainty relation via differential entropy with respect to the position and momentum, which was given by
%%%%%%%%%%%%%%%%%%%%%%%%%%%%
\begin{align}
 h(P)+h(Q) \geq \log_2(e\pi),
\label{Eq.2-55}
\end{align}
%%%%%%%%%%%%%%%%%%%%%%%%%%%%
for all possible states.   Considering a random variable $P$ characterized by a probability density $\Phi(p)$, since the differential entropy can be expressed by
%%%%%%%%%%%%%%%%%%%%%%%%%%%%
\begin{align}
h(P)=-\int^\infty_{-\infty}\Phi(p)\log_2\Phi(p)dp.
\label{Eq.2-56}
\end{align}
%%%%%%%%%%%%%%%%%%%%%%%%%%%%
 Suppose $\Lambda(p)$ belongs to the Gaussian probability distribution,  which meets
%%%%%%%%%%%%%%%%%%%%%%%%%%%%
\begin{align}
\Phi(p)=\frac1{\sqrt{2\pi\Delta(P)^2}}\exp\left(\frac{-(p-\overline{p})^2}{2\Delta(P)^2}\right).
\label{Eq.2-57}
\end{align}
%%%%%%%%%%%%%%%%%%%%%%%%%%%%
with $p$'s mean being denoted by $\bar{p}$.
 Then we can substitute Eq. (\ref{Eq.2-57}) into Eq. (\ref{Eq.2-56}) and obtain
%%%%%%%%%%%%%%%%%%%%%%%%%%%%
\begin{align}
h(P)=\log_2\sqrt{2\pi e\Delta(P)}.
\label{Eq.2-58}
\end{align}
%%%%%%%%%%%%%%%%%%%%%%%%%%%%
Owing to that the Gaussian probability distribution maximizes the differential entropy as expressed in Eq. (\ref{Eq.2-56}), thus for a general distribution we have the following formula
%%%%%%%%%%%%%%%%%%%%%%%%%%%%
\begin{align}
h(P)\leq\log_2\sqrt{2\pi e\Delta(P)}.
\label{Eq.2-59}
\end{align}
%%%%%%%%%%%%%%%%%%%%%%%%%%%%

As for arbitrary observables $P$ and $Q$ linked with position and momentum are concerned, we can obtain
%%%%%%%%%%%%%%%%%%%%%%%%%%%%
\begin{align}
\log_2\left({2\pi e\Delta(P)\Delta(Q)}\right)&=\log_2\sqrt{2\pi e\Delta(P)^2}\log_2\sqrt{2\pi e\Delta(Q)^2} \label{Eq.2-60}\\
&\geq h(P)+h(Q)\\
&\geq \log_2(e\pi).
\label{Eq.2-61}
\end{align}
%%%%%%%%%%%%%%%%%%%%%%%%%%%%
Then by combining Eqs. (\ref{Eq.2-60}) and (\ref{Eq.2-61}), we can easily deduce the earliest outcome $\Delta P\Delta Q\geq\hbar/2$ for position and momentum as mentioned before.

\subsubsection{Uncertainty relation based on Shannon entropy}

As is known, Shannon entropy plays a fundamental and key role in information theory, and quantifies the amount of information in the state of a given system in the field of classical physics. By introducing the Shannon entropy, Deutsch originally presented an uncertainty relation \cite{D.Deutsch}, written as
%%%%%%%%%%%%%%%%%%%%%%%%%%%%
\begin{align}
H{ {(Q)}}+H{(R)}\geq 2{\rm log}_2\left(\frac2{1+\sqrt {c(Q,R)}}\right),
\label{Eq.2-5}
\end{align}
%%%%%%%%%%%%%%%%%%%%%%%%%%%%
with $ H(Q) =  - \sum\nolimits_i {{p_q}\log_2 {p_q}} $ representing the Shannon entropy and ${p_q}=\mathrm{Tr}\left(| Q_q\rangle\langle Q_q|\rho\right)$ is the probability of the outcome $q$ for $Q$. The overlap $c(Q,R)=\max_{i,j}\left\{|\langle Q_i|R_j\rangle|^2\right\}$ with $\left\{ | {{Q_i}} \rangle \right\}$ and $\left\{ | {{R _j}} \rangle \right\} $ being the eigenvectors of $Q$ and $R$, respectively. Soon afterwards, Kraus \cite{K.Kraus} and Maassen and Uffink \cite{H.Maassen} together made an improvement to the result of Deutsch as
%%%%%%%%%%%%%%%%%%%%%%%%%%%%
\begin{align}
H{ {(Q)}}+H{(R)}\geq {\rm log}_2 \frac1{c(Q,R)}=:q_{MU}.
\label{Eq.2-6}
\end{align}
%%%%%%%%%%%%%%%%%%%%%%%%%%%%
It is clear to obtain that the lower bounds of Eqs. (\ref{Eq.2-5}) and (\ref{Eq.2-6}) are state independent. Korzekwa and his corporators \cite{K.Korzekwa} stated that by considering the total uncertainties, the Maassen-Uffink inequality for a qubit system can be improved as:
%%%%%%%%%%%%%%%%%%%%%%%%%%%%
\begin{align}
H{ {(Q)}}+H{(R)}\geq {\rm log}_2 \frac1{c(Q,R)}+H(\rho)\left[2+{\rm log}_2 {c(Q,R)}\right].
\label{Eq.44}
\end{align}
%%%%%%%%%%%%%%%%%%%%%%%%%%%%

\subsubsection{Uncertainty relations based on R\'{e}nyi entropy}

Stemming from the Shannon entropy, there are relatively general versions of entropies proposed by R\'{e}nyi \cite{RE} that can offer more weight to events with
 either high or low information. Owing to their inherently mathematical properties, these different types of entropies can be well applied to quantum cryptography and information theory. In general, R\'{e}nyi entropy with order $x$ is defined as
%%%%%%%%%%%%%%%%%%%%%%%%%%%%
\begin{align}
H_{x}(Q)=\frac1{1-x}{\rm log}_2\sum_q p_q^{x},
\label{Eq.4}
\end{align}
%%%%%%%%%%%%%%%%%%%%%%%%%%%%
with $x \in[0,\infty]$. For the extremity $x =1$, the R\'{e}nyi entropy recovers the Shannon entropy. In this sense, we say that the R\'{e}nyi entropy is deemed as a generalization of Shannon entropy.
{{Based on  R\'{e}nyi entropy, for generalized positive-operator-valued measure (POVM) measurements, the uncertainty relation is derived by utilizing a direct-sum majorization relation \cite{bk1}}}.
In addition, Maassen and Uffink \cite{H.Maassen} have shown that Eq. (\ref{Eq.2-6}) could be generalzied to more general case by using the R\'{e}nyi entropies. When $x,\ y\geq1/2$ and $1/x+1/y=2$ hold, one can obtain
%%%%%%%%%%%%%%%%%%%%%%%%%%%%
\begin{align}
H_{x}(Q)+H_{y}(R)\geq{\rm log}_2\frac1{c(Q,R)},
\label{Eq.5}
\end{align}
%%%%%%%%%%%%%%%%%%%%%%%%%%%%
and when $x\rightarrow \infty$ and $y\rightarrow 1/2$,
one shall attain an alternative interesting special case of Eq. (\ref{Eq.2-6}) based on the concepts of minimal and maximal entropies
%%%%%%%%%%%%%%%%%%%%%%%%%%%%
\begin{align}
H^{min}(Q)+H^{max}(R)\geq{\rm log}_2\frac1{c(Q,R)}.
\label{Eq.6}
\end{align}
%%%%%%%%%%%%%%%%%%%%%%%%%%%%
Because the minimal entropy describes the probability of accurately predicting the measurement outcome of $Q$, the above uncertain relation can be regarded as the most available relation with applications in quantum cryptography and quantum-information theory \cite{Ghasemi}.

Besides, there are also some works concentrating on the energy-time uncertainty relations \cite{Dodonov1}. In particular, Rastegin \cite{Rastegin} had derived the EUR for energy and time by means of the Pegg's approach \cite{Pegg1}.

%\subsection{Coherence uncertainty relations}
%Korzekwa et al. \cite{K.Korzekwa} presented that the total entropic uncertainty of two
%non-commuting observables can be divided into a classical part ${ C}(S,\rho)$ and an intrinsically quantum mechanical part ${ Q}(S,\rho)$.
%There exist several nontrivial features concerning classical and quantum uncertainties as follows \cite{K.Korzekwa,S.Luo}:\\
%(1) If a state $\rho$ is pure, then classical uncertainty $C(S,\rho)$ should vanish.\\
%(2) If $[\rho,S] = 0$, then the state is diagonal in the eigenbasis
%of the observable $S$ and so $Q(S,\rho)$ should vanish.\\
%(3) Classical mixing increases the classical, but not the
%quantum, uncertainty, and so $Q(S, \cdot)$ should be convex and
%$C(S, \cdot)$ should be concave in their second arguments.
%\\
%(4) $0 \leq Q(S,\rho), C(S,\rho) \leq H (S)$.\\
%(5) $Q(S, \cdot)$ and $C(S, \cdot)$ are functions of the probability
%distribution over the measurement outcomes of observable $S$
%and not of its eigenvalues.
%viz.
%\begin{align}
%H(S)= Q(S,\rho)+C(S,\rho).
%\label{Eq.6}
%\end{align}

\subsection {Majorization uncertainty relations}

There is another way to derive uncertainty relations, \emph{e.g.}, the majorization technology. This uncertainty relation, originally presented by Partovi \cite{Partovi}, was derived by the products of probabilities rather than the sums of probabilities. Subsequently, this relation was developed and generalized by Friedland \emph{et al.} \cite{Friedland} and Pucha{\l}a \emph{et al.} \cite{Pucha}. For two positive operator-valued measures (POVMs) ${\cal Q}= \{{\cal Q}^q \}_q$ and ${\cal R}=\left \{{\cal R}^r \right\}_r$, according to the general Born rule, we have these distributions $P_Q(q)=\mathrm{Tr}\left(\rho {\cal Q}^q\right)$ and $P_R(r)=\mathrm{Tr} \left(\rho {\cal R}^r \right)$ due to the measurements ${\cal Q}$ and ${\cal R}$ on $\rho$, respectively. We now denote $P^{\downarrow}_{Q}$ and $P^{\downarrow}_{R}$ as the corresponding reordered vectors in order to rank the probabilities from largest to smallest.

In order to seek a vector that majorizes the tensor product of a pair of probability vectors $P^{\downarrow}_{Q}$ and $P^{\downarrow}_{R}$, {\it i.e.}, we set up a probability distribution $\mu=\left\{\mu(1),\mu(2),\ldots,\mu(Q\|R)\right\}$ such that
%%%%%%%%%%%%%%%%%%%%%%%%%%%%
\begin{equation} \label{Eq.2-8}
P^{\downarrow}_{Q} \times  P^{\downarrow}_{R} \prec \mu \ \ (\forall \rho),
\end{equation}
%%%%%%%%%%%%%%%%%%%%%%%%%%%%
which offers the bound regarding how spread-out the product distribution $P^{\downarrow}_{Q} \times  P^{\downarrow}_{R}$ should be. To seek the probability distribution $\mu$ matching Eq. (\ref{Eq.2-8}), we can take into account the largest probability with regard to a product distribution in Eq. (\ref{Eq.2-8}), shown as
%%%%%%%%%%%%%%%%%%%%%%%%%%%%
\begin{equation} \label{Eq.9}
p_1=P^{\downarrow}_{Q} \cdot P^{\downarrow}_{R} =p_\mathrm{guess}(Q)\cdot p_\mathrm{guess}(R).
\end{equation}
%%%%%%%%%%%%%%%%%%%%%%%%%%%%
It is well known that $p_1$ will be always left away from 1 with respect to two incompatible measurements, on account that the two measurements cannot get deterministic outcomes at the same time. From Deutsch's result \cite{D.Deutsch}, we have $p_\mathrm{guess}(Q)  p_\mathrm{guess}(R)\leq b^2$, which will yield
%%%%%%%%%%%%%%%%%%%%%%%%%%%%
\begin{equation} \label{Eq.10}
p_1=p_\mathrm{guess}(Q)  p_\mathrm{guess}(R)\leq b^2=:\mu_1,
\end{equation}
%%%%%%%%%%%%%%%%%%%%%%%%%%%%
as to the orthonormal bases ${\cal Q}$ and ${\cal R}$, with $b=\frac {1}{2}[1+\sqrt {c}]$. Thereby, one can easily see that the vector $\mu^1= \{\mu_1,1-\mu_1,0, \ldots,0\}$ meets Eq. (\ref{Eq.2-8}) and factually makes up a simple and nontrivial uncertainty relation.

Besides, there are two important works in which an effective methodology was proposed to build a sequence of vectors $\{ \mu^m\}_{m=1}^{|Q|-1}$
of the form \cite{Friedland,Pucha}
%%%%%%%%%%%%%%%%%%%%%%%%%%%%
\begin{equation} \label{Eq.11}
\mu^m=\left\{\mu_1,\mu_2-\mu_1,\ldots,1-\mu_{m-1},0,\ldots,0 \right\},
\end{equation}
%%%%%%%%%%%%%%%%%%%%%%%%%%%%
with $\mu^{m} \prec \mu^{m-1}$ that complies with Eq. (\ref{Eq.2-8}), and this will result in a tight uncertainty relation. Certainly,
the expressions of $\mu_{m}$ can be derived according to an improvement and also will be gradually more and more difficult with the increasing $m$.

On account R\'{e}nyi entropy is Schur concave and additive, R\'{e}nyi-entropy-based uncertainty relation follows straightly from the aforementioned majorization relations. With this in mind, we have
%%%%%%%%%%%%%%%%%%%%%%%%%%%%
\begin{equation} \label{Eq.13}
P^{\downarrow}_{Q} \cdot P^{\downarrow}_{R}\prec \mu \Rightarrow H_x({\cal Q})+H_x({\cal R})\geq H_{x}({\cal V}),
\end{equation}
%%%%%%%%%%%%%%%%%%%%%%%%%%%%
where ${\cal V}$ denotes a random variable and its distribution is in accordance with the law $\mu$. The uncertainty relation has a diverse flavor compared with the Maassen-Uffink relation in Eq. (\ref{Eq.2-6}) as it gives a lower bound for the summation of the R\'{e}nyi entropies with the same parameters.
With respect to a particular case of $x \rightarrow \infty$, one can naturally recover the uncertainty relation presented by Deutsch \cite{D.Deutsch},
%%%%%%%%%%%%%%%%%%%%%%%%%%%%
\begin{equation} \label{Eq.2-14}
H(Q)+H(R)\geq H^{min}(Q)+H^{min}(R)\geq {\log}_2 \frac 1{b^2}.
\end{equation}
%%%%%%%%%%%%%%%%%%%%%%%%%%%%
For the first inequality in Eq. (\ref{Eq.2-14}), it was derived from the monotonicity of the R\'{e}nyi entropies being relevant with the parameter $x$. Note that, in terms of Eq. (\ref{Eq.13}), one can easily obtain
%%%%%%%%%%%%%%%%%%%%%%%%%%%%
\begin{equation} \label{Eq.15}
H(Q)+H(R)\geq H_\mathrm{bin} \left(b^2\right)=:q_{\rm majorization},
\end{equation}
if $x=1$ is chosen. Here, $H_\mathrm{bin}(\theta)=-\theta\log_2 \theta-(1-\theta)\log_2(1-\theta)$ denotes the binary Shannon entropy function.
%%%%%%%%%%%%%%%%%%%%%%%%%%%%

\section{Quantum-memory-assisted EUR} \label{sec:3}
In Sec. 2, we reviewed recent progresses of the EURs for which
the observer Bob can only access to the classical information,
that is, the information about the preparation of the particle to be
measured. From a practical point of view, it is also appealing to
further examine the situation for which Bob can access to the
quantum information. More specifically, the particle $A$ to be
measured by Alice is quantum correlated with another particle $B$
(served as the quantum memory) holds by Bob. For this case, Bob is
equipped to use the quantum information communicated between $A$ and
$B$. As such, his guessing probability about Alice's measurement
outcomes may be enhanced. A prior attempt along this line was
completed by Renes and Boileau \cite{Renes}. They showed that for
two complementary observables $X$ and $Z$, we have
%%%%%%%%%%%%%%%%%%%%
\begin{equation}\label{eq3-1}
  S(X|B)+S(Z|B)\geq \log_2 d + S(A|B),
\end{equation}
%%%%%%%%%%%%%%%%%%%%
and
\begin{equation}\label{eq3-2}
 S(X|B)+S(Z|E)\geq \log_2 d,
\end{equation}
%%%%%%%%%%%%%%%%%%%%
with $d$ being the $A$'s dimension, and $E$ being the system
possessed by an eavesdropper.

Later, Berta \emph{et al.} \cite{Berta} generalized the above EUR to
arbitrary two observables. They considered an imaginary
"uncertainty game" between two players (Alice and Bob) who agreed
on two measurements $Q$ and $R$ in advance. Bob entangles his
particle $B$ with another particle $A$ that he sends to Alice. Alice
then carries on anyone of the measurements chosen at random on her
particle and only broadcasts her measurement choice to Bob. Bob's task
is to guess as precise as possible Alice's outcome by measuring
his particle $B$ with the help of the received classical information
({\it i.e.}, Alice's choice of measurement). By taking $S(Q|B)$ as Bob's
uncertainty about Alice's measurement outcome of the observable $Q$,
and similarly for $S(R|B)$, Berta \emph{et al.} \cite{Berta} proved
strictly the following quantum-memory-assisted EURs
%%%%%%%%%%%%%%%%%%%%
\begin{align}\label{eq3-3}
 S(Q|B)+S(R|B)&\geq \log_2 \frac{1}{c} + S(A|B), \end{align}
 and
 \begin{align}\label{eq3-4}
 S(Q|B)+S(R|E)&\geq \log_2 \frac{1}{c},
\end{align}
%%%%%%%%%%%%%%%%%%%%
where $S(A|B)=S(AB)-S(B)$ is denoted as the conditional entropy of the
premeasurement state $\rho_{AB}$, while $S(Q|B)$ denotes the conditional
entropy of the postmeasurement state
%%%%%%%%%%%%%%%%%%%%
\begin{equation}\label{eq3-5}
 \rho_{QB}= \sum_i \left(\Pi_i^Q \otimes \mathds{1}_B\right)\rho_{AB}\left(\Pi_i^Q\otimes \mathds{1}_B\right),
\end{equation}
%%%%%%%%%%%%%%%%%%%%
where $\Pi_i^Q=|\psi_i^Q\rangle\langle\psi_i^Q|$ are the measurement
operators on $\mathcal{H}_A$ with $\{|\psi_i^Q\rangle\}$ being the
eigenvectors of the observable $Q$, $\mathds{1}_B$ is the identity
operator on $\mathcal{H}_B$, and $S(R|B)$ is similarly defined.
Moreover, the parameter $c=\max_{ij} \{c_{ij}\}$ in Eqs. \eqref{eq3-3} and \eqref{eq3-4}
 measures the complementarity of $Q$ and $R$, with
%%%%%%%%%%%%%%%%%%%%
\begin{equation}\label{eq3-6}
 c_{ij}= |\langle \psi_i^Q|\psi_j^R\rangle|^2.
\end{equation}
%%%%%%%%%%%%%%%%%%%%

Compared with Eq. (\ref{Eq.2-6}) in Sec. 2, one can see that
the uncertainty bound given on the right-hand side (RHS) of Eq.
\eqref{eq3-3} will be reduced for the negative conditional entropy
$S(A|B)$. In particular, for the case of the observables $Q$ and $R$
being complementary such that $c=1/d$, and the particles $A$ and $B$
being maximally entangled for which $S(A|B)= -\log_2 d$, the term on
the RHS of Eq. \eqref{eq3-3} (we call it Berta \emph{et al.}'s
uncertainty bound hereafter) is reduced to zero. As a result, Bob
will successfully in predicting Alice's measurement outcomes of both
$Q$ and $R$ precisely.

Experimentally, the quantum-memory-assisted EUR of Eq. \eqref{eq3-3}
has been demonstrated in all-optical set-ups \cite{Nphys1,Nphys2}.
Moreover, a proposal for testing it in the nitrogen-vacancy center
in diamond is also presented \cite{APL}.

\subsection{Improved lower bounds of the EUR}
By considering quantum correlations between the particles $A$ and
$B$, one can derive tighter uncertainty bounds than those of Eqs.
\eqref{eq3-3} and \eqref{eq3-4}. Pati \emph{et al.} \cite{Pati}
considered such a problem. Starting from the concept of quantum
discord \cite{QD}, they proved that the uncertainty bound of Eq.
\eqref{eq3-3} can be tightened as
%%%%%%%%%%%%%%%%%%%%
\begin{equation}\label{eq3a-1}
  S(Q|B)+S(R|B)\geq \log_2 \frac{1}{c}+S(A|B)+\max\{0,-\delta_2\},
\end{equation}
%%%%%%%%%%%%%%%%%%%%
where $\delta_2= J(B|A)-D(B|A)$, $J(B|A)$ is the classical correlation and
$D(B|A)$ is the quantum discord \cite{QD}. They are
given by
%%%%%%%%%%%%%%%%%%%%%%%%%%%
\begin{equation}\label{eq3a-2}
 \begin{split}
  & J(B|A)= S(\rho_B)-\min_{\{E_k^A\}} S(B|\{E_k^A\}), \\
  & D(B|A)= I(\rho_{AB}) -J(B|A),\\
 \end{split}
\end{equation}
%%%%%%%%%%%%%%%%%%%%%%%%%%%
where $I(\rho_{AB})=S(\rho_A)+S(\rho_B)-S(\rho_{AB})$ denotes the quantum
mutual information of $\rho_{AB}$, and
%%%%%%%%%%%%%%%%%%%%%%%%%%%
\begin{equation}\label{eq3a-3}
 S(B|\{E_k^A\})=\sum_{k}p_{k}S(\rho_{B|E_k^A})
\end{equation}
%%%%%%%%%%%%%%%%%%%%%%%%%%%
with $\rho_{B|E_k^A}= \mathrm{Tr}_{A}(E_k^A\rho_{AB})/p_k$ representing the
postmeasurement state of the POVM $E_k^A$, and $p_k= \mathrm{Tr}
(E_k^A\rho_{AB})$ is the probability of the outcome $k$.

The key point for proving Eq. \eqref{eq3a-1} is $S(X|B)=
{{S(X)}}-I(\rho_{XB})$ ($X=Q$ or $R$), $I(\rho_{XB}) \leq J(B|A)$,
and the EUR of Eq. \eqref{Eq.2-6} in Sec. 2. It indicates that
whenever the quantum discord exceeds the classical correlation,
the uncertainty bound in Eq. \eqref{eq3a-1} tightens Berta \emph{et
al.}'s uncertainty bound. By taking into account the purification
$|\Psi\rangle_{ABC}$ of $\rho_{AB}$, one can also show that the
correlation discrepancy of $J(B|A)$ and $D(B|A)$ equals the monogamy
score \cite{pra014105,score}
%%%%%%%%%%%%%%%%%%%%%%%%%%%
\begin{equation}\label{eq3a-4}
 \delta_D= D(BC|A)-D(B|A)-D(C|A),
\end{equation}
%%%%%%%%%%%%%%%%%%%%%%%%%%%
hence Berta \emph{et al.}'s uncertainty bound is improved only if
the purification $|\Psi\rangle_{ABC}$ of $\rho_{AB}$ violates the
monogamy inequality $ D(BC|A)\geq D(B|A)+D(C|A)$.

Similarly, the uncertainty bound of Eq. \eqref{eq3-4} can be
tightened as \cite{Pati}
%%%%%%%%%%%%%%%%%%%%
\begin{equation}\label{eq3a-5}
 S(Q|B)+S(R|E)\geq  \log_2 \frac{1}{c}+\max\{0,-\delta'_2\},
\end{equation}
%%%%%%%%%%%%%%%%%%%%
where $\delta'_2= J(B|A)-D(BE'|A)$, with $D(BE'|A)$ being the
quantum discord between $A$ and $BE'$, and $E'$ is the purifying
system of $ABE$, {\it i.e.}, $\rho_{ABE}= \mathrm{Tr}_{E'}
(|\Psi\rangle_{ABEE'}\langle \Psi|)$.

Coles and Piani \cite{bound1} also explored the improved uncertainty
bound of the EUR. By denoting $c_2$ the second largest value of $\{c_{ij}\}$,
they first proved that
%%%%%%%%%%%%%%%%%%%%
\begin{equation}\label{eq3a-6}
  S(Q|B)+S(R|B)\geq  \log_2 \frac{1}{c}+\frac{1-\sqrt{c}}{2}\log_2\frac{c}{c_2} + S(A|B),
\end{equation}
%%%%%%%%%%%%%%%%%%%%
and further proved the following tight uncertainty bound
%%%%%%%%%%%%%%%%%%%%
\begin{equation}\label{eq3a-7}
 S(Q|B)+S(R|B)\geq  q(\rho_A) + S(A|B),
\end{equation}
%%%%%%%%%%%%%%%%%%%%
where $q(\rho_A)=\max\{q(\rho_A,Q,R),q(\rho_A,R,Q)\}$, and
%%%%%%%%%%%%%%%%%%%%
\begin{equation}\label{eq3a-8}
 \begin{split}
  & q(\rho_A,Q,R)= \sum_j p_j^Q \log_2 \frac{1}{\max_k c_{jk} },\\
  & q(\rho_A,R,Q)= \sum_k p_k^R \log_2 \frac{1}{\max_j c_{jk} },
 \end{split}
\end{equation}
%%%%%%%%%%%%%%%%%%%%
where $p_j^Q= \mathrm{Tr} (\Pi_i^Q \rho_{A})$ is the measurement
outcome probability distribution of $Q$, and likewise for $p_k^R$.
It is straightforward to see that this uncertainty bound is the same
to Berta \emph{et al.}'s uncertainty bound for $d=2$, and it may be
tighter than Berta \emph{et al.}'s uncertainty bound for $d\geq 3$.

One can also obtain a relative weak bound by minimizing $q(\rho_A)$
over the full set of $\rho_A$, {\it i.e.}, $q=\min_{\rho_A} q(\rho_A)$.
Coles and Piani \cite{bound1} proved that this minimization can be
achieved via the following procedure:
%%%%%%%%%%%%%%%%%%%%
\begin{equation}\label{eq3a-9}
 q= \max_{0\leq p \leq 1} \lambda_{\min}[\Delta(p)],
\end{equation}
%%%%%%%%%%%%%%%%%%%%
where $\lambda_{\min}[\Delta(p)]$ represents the minimum eigenvalue of
the matrix $\Delta(p)=p\Delta_{QR}+(1-p)\Delta_{RQ}$, with
%%%%%%%%%%%%%%%%%%%%
\begin{equation}\label{eq3a-10}
 \begin{aligned}
  & \Delta_{QR}= \sum_j \log_2 (1/\max_k c_{jk} ) |\psi_j^Q\rangle\langle \psi_j^Q|, \\
  & \Delta_{RQ}= \sum_k \log_2 (1/\max_j c_{jk}) |\psi_k^R\rangle\langle \psi_k^R|.
 \end{aligned}
\end{equation}
%%%%%%%%%%%%%%%%%%%%

Of course, in a similar manner to prove Eq. \eqref{eq3a-1}, the
uncertainty bounds of Eqs. \eqref{eq3a-6} and \eqref{eq3a-7} can be
further tightened by adding the additional term $\max\{0, -\delta_2
\}$ to the RHS of them.

{{Adabi \emph{et al.} \cite{bound2} and Haseli \emph{et al.} \cite{hs22} }}discussed the improved uncertainty
bound of the EUR from the perspective of Holevo quantity and mutual
information. To be explicit, they showed that
%%%%%%%%%%%%%%%%%%%%
\begin{equation}\label{eq3a-11}
  S(Q|B)+S(R|B)\geq \log_2 \frac{1}{c}+S(A|B) +\max\{0, \chi_2\},
\end{equation}
%%%%%%%%%%%%%%%%%%%%
where
%%%%%%%%%%%%%%%%%%%%
\begin{equation}\label{eq3a-12}
 \chi_2= I(\rho_{AB})-I(\rho_{QB})-I(\rho_{RB}),
\end{equation}
%%%%%%%%%%%%%%%%%%%%
with $I(\rho_{QB})$ measuring Bob's accessible information with regard to
Alice's measurement $Q$, and likewise for $I(\rho_{RB})$. So
whenever the quantum mutual information $I(\rho_{AB})$ is larger than
the sum of Bob's accessible information, the uncertainty bound given
in the RHS of Eq. \eqref{eq3a-11} will be tighter than Berta \emph{et al.}'s
uncertainty bound. For pure state, as $\delta_2=\chi_2=0$, this
uncertainty bound coincides with those given in Eqs. \eqref{eq3-3}
and \eqref{eq3a-1}. Adabi \emph{et al.} \cite{bound2} also showed
that for Werner states it coincides with that of Eq. \eqref{eq3a-1},
while for Bell-diagonal states and two-qubit \textit{X} states, it
turns out to be tighter than those of Eqs. \eqref{eq3-3} and
\eqref{eq3a-1}.

\subsection{Generalized quantum-memory-assisted EURs}
The quantum-memory-assisted EUR proved by Berta \emph{et al.}
\cite{Berta} applies to the case of two observables, but it can also be
generalized to the general case of multiple measurement settings.
Along this line, several progresses have been made  recently, and it is hoped to bring further understanding about
uncertainty principle which differentiates quantum mechanics from
the classical world.

For the imaginary "uncertainty game" constructed by Berta
\emph{et al.} with however the two measurements $\{Q,R\}$ being
replaced by the $N$ measurements $\{M_i\}_{i=1}^N$, Liu \emph{et
al.} \cite{LMF} obtained the following bound as to Bob's
uncertainty regarding Alice's measurement results
%%%%%%%%%%%%%%%%%%%%
\begin{align}\label{eq3b-1}
 &\sum_{i=1}^N S(M_i|B) \geq \log_2 \frac{1}{b}+ (N-1)S(A|B),\\
 &b= \max_{i_N}\left\{\sum_{i_2 \sim i_{N-1}}\max_{i_1}\big[c(\psi_{i_1}^1,\psi_{i_2}^2)\big]
    \prod_{m=2}^{N-1}c( \psi_{i_m}^m, \psi_{i_{m+1}}^{m+1}) \right \}.
\end{align}
%%%%%%%%%%%%%%%%%%%%
and by defining
%%%%%%%%%%%%%%%%%%%%
\begin{equation}\label{eq3b-2}
 c(\psi_{i_m}^m, \psi_{i_n}^n)=\max_{i_m i_n}|\langle \psi_{i_m}^m|\psi_{i_n}^n\rangle|^2,
\end{equation}
%%%%%%%%%%%%%%%%%%%%
with $\{|\psi_{i_m}^m\rangle\}$ denoting the eigenvectors of $M_m$,
the parameter $b$ can be obtained as
%%%%%%%%%%%%%%%%%%%%
\begin{equation}\label{eq3b-3}
 b= \max_{i_N}\left\{\sum_{i_2 \sim i_{N-1}}\max_{i_1}\big[c(\psi_{i_1}^1,\psi_{i_2}^2)\big]
    \prod_{m=2}^{N-1}c( \psi_{i_m}^m, \psi_{i_{m+1}}^{m+1}) \right \}.
\end{equation}
%%%%%%%%%%%%%%%%%%%%
Clearly, when $N=2$, $b$ reduces to that of $c$ given in Eq.
\eqref{eq3-6}, that is, the uncertain relation of Eq. \eqref{eq3b-1}
covers that of Berta \emph{et al.} as a special case.

By denoting $\varepsilon$ a new order of the measurements $\{M_i\}$
and $\left\{|\varepsilon_{i_m}^m\rangle\right\}$ the corresponding eigenvectors
of $M_m$ in the $\varepsilon$ order, Zhang \emph{et al.} \cite{Yucs}
obtained a tighter lower uncertainty bound than Eq. \eqref{eq3b-1}.
Their result is
%%%%%%%%%%%%%%%%%%%%
\begin{equation}\label{eq3b-4}
 \sum_{i=1}^N S(M_i|B) \geq \max_\varepsilon \{\ell_\varepsilon\}+(N-1)S(A|B),
\end{equation}
%%%%%%%%%%%%%%%%%%%%
where
%%%%%%%%%%%%%%%%%%%%
\begin{equation}\label{eq3b-5}
 \ell_\varepsilon= -\sum_{i_N} p_{\varepsilon_{i_N}^N} \log_2
                   \sum_{i_k, N\geq  k>1} \max_{i_1} \prod_{n=1}^{N-1}
                   \left|\langle \varepsilon_{i_n}^n|\varepsilon_{i_{n+1}}^{n+1}\rangle\right|^2,
\end{equation}
%%%%%%%%%%%%%%%%%%%%
with $p_{\varepsilon_{i_N}^N}= \mathrm{Tr} (|\varepsilon_{i_N}^N\rangle
\langle\varepsilon_{i_N}^N|\otimes\mathds{1}_B)\rho_{AB}$. For the case
of $N=2$, this bound reduces to that of Eq. \eqref{eq3a-7}.

One can also tighten the uncertainty bound of Eq. \eqref{eq3b-1} by
using Bob's accessible information in a manner similar to Ref.
\cite{bound2}. Dolatkhah \emph{et al.} \cite{qip1} showed that
%%%%%%%%%%%%%%%%%%%%
\begin{equation}\label{eq3b-6}
  \sum_{i=1}^N S(M_i|B)\geq \log_2 \frac{1}{b}+ (N-1)S(A|B)
                       +\max\{0, \chi_N\},
\end{equation}
%%%%%%%%%%%%%%%%%%%%
with $\chi_N=(N-1)I(\rho_{AB})-\sum_{i=1}^N I(\rho_{M_{i}B})$. This
uncertainty bound is stronger than Eq. \eqref{eq3b-5} as $J(B|A)\leq
I(\rho_{M_i B})$ ($\forall M_i$) \cite{QD}.

Moreover, by using the similar methodology as proving Eq.
\eqref{eq3a-1} and the EUR of multiple measurements without a quantum memory \cite{LMF}
%%%%%%%%%%%%%%%%%%%%
\begin{equation}\label{eq3b-7}
 \sum_{i=1}^N S(M_i) \geq  \log_2 \frac{1}{b}+ (N-1)S(\rho_A),
\end{equation}
%%%%%%%%%%%%%%%%%%%%
one can show immediately that
%%%%%%%%%%%%%%%%%%%%%%%%%%%
\begin{equation}\label{eq3b-8}
 \begin{split}
  \sum_{i=1}^N S(M_i|B)=& \sum_{i=1}^N {{S(M_i)}}-\sum_{i=1}^N I(\rho_{M_i B}) \\
                       \geq &  \sum_{i=1}^N {{S(M_i)}} - NJ(B|A) \\
                       \geq &  \log_2 \frac{1}{b}+ (N-1)S(\rho_A)- NJ(B|A) \\
                       =& \log_2 \frac{1}{b}+ (N-1)S(A|B)\\
                       &+ (N-1)D(B|A)-J(B|A),
 \end{split}
\end{equation}
%%%%%%%%%%%%%%%%%%%%%%%%%%%
then one can obtain a tighter lower uncertainty bound than that of
Eq. \eqref{eq3b-1} as
%%%%%%%%%%%%%%%%%%%%
\begin{equation}\label{eq3b-9}
  \sum_{i=1}^N S(M_i|B)\geq \log_2 \frac{1}{b}+ (N-1)S(A|B)+ \max\{0,-\delta_N\}.
\end{equation}
%%%%%%%%%%%%%%%%%%%%
where $\delta_N= J(B|A)-(N-1)D(B|A)$.

Hu and Fan \cite{pra022314} investigated the quantum-memory-assisted
EUR from another perspective. They generalized the "uncertainty
game" of Berta \emph{et al.} \cite{Berta} to the scenario of $N$
players who share the state $\rho_{AB_1B_2 \ldots B_{N-1}}$, the
explicit form of which is known to all the players other than Alice.
The tasks for the players $B_1B_2 \ldots B_{N-1}$ (communications
among them are forbidden) are to predict Alice's measurement
outcomes on particle $A$. Relying on the strong subadditivity of the
von Neumann entropy and the subadditivity of the conditional entropy
\cite{Nielsen}, it was shown that \cite{pra022314}
%%%%%%%%%%%%%%%%%%%%
\begin{equation}\label{eq3b-10}
  \sum_{i=1}^{N-1} S(A|B_i)\geq 0,
\end{equation}
%%%%%%%%%%%%%%%%%%%%
which indicates that for this scenario, Alice's measurement
outcomes about particle $A$ cannot be predicted correctly by the
players $B_1B_2 \ldots B_{N-1}$ simultaneously. This may be
recognized as another kind of uncertainty relation.

\subsection{Interpreting the quantum-memory-assisted EUR}
While $-S(A|B)$ is a tight lower bound of the one-way distillable
entanglement, the uncertainty bound of Eq. \eqref{eq3-2} is not a
monotonic function of the amount of entanglement between the two
particles. This stimulates further research aimed at revealing the
intrinsic connections between the reduced entropic uncertainty and
the quantum correlation of the measured particle and the quantum
memory.

By considering the purification $|\Psi\rangle_{ABC}$ of $\rho_{AB}$
shared between Alice and Bob, {\it i.e.}, $\rho_{AB}=\mathrm{Tr}_C (|\Psi
\rangle_{ABC}\langle \Psi|)$, we compared amount of quantum
correlations between the parties $AB$ and $AC$ \cite{pra022314}.
First, for quantum correlations measured by the entanglement of
formation \cite{EoF1,EoF2} and quantum discord \cite{QD}, it was
found that whenever the uncertainty bound of Berta \emph{et al.} is
reduced compared with that without a quantum memory, we always have
\cite{pra022314}
%%%%%%%%%%%%%%%%%%%%
\begin{equation}\label{eq3c-1}
  E_f(\rho_{AB})>E_f(\rho_{AC}),~
  D(B|A)>D(C|A),
\end{equation}
%%%%%%%%%%%%%%%%%%%%
and to prove the above inequalities, one can adopt the Koashi-Winter
equality \cite{Koashi}. Second, if one considers the quantum
correlations measured by the one-way unlocalizable quantum entanglement
\cite{ulqe} and one-way unlocalizable quantum discord \cite{ulqd},
then by using the Buscemi-Gour-Kim equality \cite{ulqe}, one can
show that \cite{pra022314}
%%%%%%%%%%%%%%%%%%%%
\begin{equation}\label{eq3c-2}
  E_u^\leftarrow(\rho_{BA})>E_u^\leftarrow(\rho_{CA}),~
  \delta_u^\leftarrow(\rho_{BA})>\delta_u^\leftarrow(\rho_{CA}),
\end{equation}
%%%%%%%%%%%%%%%%%%%%
when Berta \emph{et al.}'s uncertainty bound is tighter than the
bound obtained without a quantum memory. All these observations show
that the presence of a quantum memory helps improving prediction
precision of Alice's outcomes only when it is quantum correlated
with the measured particle $A$ in a way stronger than its
correlation with the purifying system $C$.

Moreover, it has also been shown that \cite{pra022314}
%%%%%%%%%%%%%%%%%%%%
\begin{equation}\label{eq3c-3}
 S(A|B)=D(C|A)-D(B|A),
\end{equation}
%%%%%%%%%%%%%%%%%%%%
therefore Berta \emph{et al.}'s uncertainty bound is dependent
quantitatively of the competition between quantum discords $D(C|A)$
and $D(B|A)$, and it is decreased monotonically with the decrease of
$D(C|A)-D(B|A)$.

While the strength of quantum correlations constraint the prediction
precision of the measurement outcomes in the "uncertainty game",
from another point of view, the EUR also imposes constraints on the
amount of quantum correlations in the bipartite state of the measured
particle and the quantum memory, which might be employed to derive bounds
on quantum correlations.

When the quantum correlation is measured by quantum discord, it was
shown in Ref. \cite{pra014105} that one may get the following
tight upper bound
%%%%%%%%%%%%%%%%%%%%%%%%%%%
\begin{eqnarray}\label{eq3c-4}
 D(B|A) \leq \min \left\{S(\rho_A),I(\rho_{AB}), \frac{1}{2}\left(\delta_T+I(\rho_{AB})\right)\right\},
\end{eqnarray}
%%%%%%%%%%%%%%%%%%%%%%%%%%%
where
%%%%%%%%%%%%%%%%%%%%%%%%%%%
\begin{eqnarray}\label{eq3c-5}
 \delta_T= S(Q|B)+S(R|B)-\log_2 \frac{1}{c}-S(A|B),
\end{eqnarray}
%%%%%%%%%%%%%%%%%%%%%%%%%%%
which immediately recovers the result of Eq. \eqref{eq3a-1}, and the first
two terms on the RHS of Eq. \eqref{eq3c-5} can be obtained
experimentally based on the projective measurements on particle $A$
and quantum tomography on particle $B$.

By further using the inequalities $S(Q|B) \leq S(Q|Q)$ ({\it i.e.},
the projective measurements does not decrease entropy, and $S(Q|Q)$
is the conditional entropy of $\rho_{QQ}= \Pi^Q \otimes \Pi^Q
(\rho_{AB})$) and $H(X|B) \leq h(p_X) + p_X \log_2(d-1)$ ({\it i.e.},
the Fano's inequality, where $p_X$ is the probability of different
outcomes of measurements $X$ on $A$ and $B$) \cite{Nielsen}, two a
bit weaker but experimentally more accessible bounds of quantum
discord were derived as \cite{pra014105}
%%%%%%%%%%%%%%%%%%%%%%%%%%%
\begin{eqnarray}\label{eq3c-6}
 D(B|A) \leq \min \left\{S(\rho_A),I(\rho_{AB}),\frac{1}{2}[\delta_{\rm \alpha}+I(\rho_{AB})]\right \},
\end{eqnarray}
%%%%%%%%%%%%%%%%%%%%%%%%%%%
where for $\alpha=M$ the parameter $\delta_M$ can be obtained
directly by replacing the first two terms on the RHS of Eq.
\eqref{eq3c-5} with $S(Q|Q)+S(R|R)$, and for $\alpha=F$ the
parameter $\delta_F$ can be obtained by replacing the first two
terms on the RHS of Eq. \eqref{eq3c-5} with $h(p_Q)+h(p_R)+(p_Q+p_R)
\log_2(d-1)$.

The quantum-memory-assisted EUR is also intimately related to
the monogamy properties of quantum discord, e.g., for the purification
$|\Psi\rangle_{ABC}$ of $\rho_{AB}$, it was shown that
\cite{pra014105}
%%%%%%%%%%%%%%%%%%%%%%%%%%%
\begin{eqnarray}\label{eq3c-7}
 \begin{split}
  & D(B|A)+ D(C|A) \leq D(BC|A)+ \delta_T, \\
  & D(A|B)+ D(A|C) \leq D(A|BC)+ \bar{\delta}_T,
 \end{split}
\end{eqnarray}
%%%%%%%%%%%%%%%%%%%%%%%%%%%
where $\delta_T$ is given in Eq. \eqref{eq3c-5}, and
$\bar{\delta}_T= (\delta_T^{BA}+\delta_T^{CA})/2$, with
%%%%%%%%%%%%%%%%%%%%%%%%%%%
\begin{eqnarray}\label{eq3c-8}
 \begin{split}
  \delta_T^{BA}=S(Q_B|A)+S(R_B|A)-\log_2 \frac{1}{c}-S(B|A), \\
  \delta_T^{CA}=S(Q_C|A)+S(R_C|A)-\log_2 \frac{1}{c}-S(C|A).
 \end{split}
\end{eqnarray}
%%%%%%%%%%%%%%%%%%%%%%%%%%%
Thus even if quantum discord is not monogamous, it still cannot be
freely shared among the three parties.

\subsection{Linking the EURs to quantum coherence}
In the presence of a quantum memory, Bob's uncertainty with respect to
Alice's measurement outcomes is characterized by the conditional von
Neumann entropy of the one-sided projective measurements on $A$, and
this entropy plays an important role in the resource theory of quantum
coherence (see Ref. \cite{PhysRep} and references therein). In
particular, it was shown that the quantum discord of a bipartite
state can be interpreted from the perspective of quantum coherence
\cite{Hupra}. Therefore, one can also explore the EURs along this
line, and this may shed some new light to the essence of EURs from a
new aspect.

Korzekwa \emph{et al.} \cite{K.Korzekwa} considered the above problem by
decomposing the total uncertainty of an observable $O$ into the quantum
and classical components, {\it  i.e.}, $H_O(\rho)= Q(O,\rho)+ C(O,\rho)$. They
defined the quantum uncertainty as $Q(O,\rho)= S(\rho\|
\rho_{\Delta_O(\rho)})$, where $\Delta_O(\rho)$ denotes the full dephasing
of $\rho$ in the reference basis spanned by the eigenbasis of $O$.
Remarkably, as far as classical and quantum uncertainties are concerned,
several compelling features they should obey are as follows:

(a) When a quantum system is prepared in a pure state $\rho$, then classical
uncertainty $C(O,\rho)$ would be zero-valued.

(b) As $[\rho,O] = 0$ holds,  $\rho$ becomes diagonal in the
basis expanded by the eigenbases of $O$, which will result in the vanishing
quantum uncertainty $Q(O,\rho)$.

(c) Classical mixing increases the classical uncertainties rather than
the quantum parts, as a result, $Q(O, \cdot)$ ought to be convex and $C(O, \cdot)$
concave in these arguments.

(d) $Q(O,\rho)\geq 0$ and $C(O,\rho) \leq H (\rho)$.

(e) $Q(O, \cdot)$ and $C(O, \cdot)$ in essence are functions of the probability
distribution over the measurement's results of observable $O$ rather than  the corresponding
eigenvalues.

From the above requirements, one can see that $Q(O,\rho)$ is just the
relative entropy of coherence $C_{r}(Q,\rho)$ defined in Ref. \cite{coher}.
As such, the corresponding uncertainty relations can be termed as
uncertainty relations of quantum coherence. Based on this decomposition,
one can show that
%%%%%%%%%%%%%%%%%%%%%%%%%%%
\begin{equation} \label{eq3d-1}
 C_r(Q,\rho)+C_r(R,\rho)\geq \log_2 \frac{1}{c}-S(\rho),
\end{equation}
%%%%%%%%%%%%%%%%%%%%%%%%%%%

Yuan \emph{et al.} \cite{eurqc2} also studied such a problem. Their
results are as follows:
%%%%%%%%%%%%%%%%%%%%%%%%%%
\begin{equation} \label{eq3d-2}
 \begin{aligned}
  & C_r(Q,\rho)+C_r(R,\rho)\geq  H_\mathrm{bin}\left(1+\frac{\sqrt{P'}(2\sqrt{c}-1)}{2} \right)-S(\rho),\\
  & C_{l_1}(Q,\rho)+C_{l_1}(R,\rho)\geq 2\sqrt{P'c(1-c)}, \\
  & R_I(Q,\rho)+R_I(R,\rho)\geq  H_\mathrm{bin}\left(\frac{1+\sqrt{1-4P'(\sqrt{c}-c)}}{2}\right),
 \end{aligned}
\end{equation}
%%%%%%%%%%%%%%%%%%%%%%%%%%
where $P'=2\mathrm{Tr}\rho^2-1$, $C_{l_1}(Q,\rho)$ is the $l_1$ norm of
coherence {{\cite{coher,Malvezzi22,he22,HML11}}}, and $R_I(Q,\rho)$ is the coherence of
formation \cite{cof1,cof2}.

Singh and coauthors \cite{eurqc3} considered instead the uncertainty
relation of quantum coherence in the presence of quantum memory.
When using the basis $\mathcal{B}_X=\{|\psi_k^X\rangle\otimes
|\varphi_l^B \rangle\}$ ($X=Q$ or $R$), where $|\varphi_l^B\rangle$
is the eigenstate of $\rho_B=\mathrm{Tr}_A \rho_{AB}$, it can be derived
from \eqref{eq3-3} that
%%%%%%%%%%%%%%%%%%%%%%%%%%%
\begin{equation}\label{eq3d-3}
 C_r(\mathcal{B}_Q,\rho_{AB})+C_r(\mathcal{B}_R,\rho_{AB})\geq \log_2\frac{1}{c}-S(A|B),
\end{equation}
%%%%%%%%%%%%%%%%%%%%%%%%%%%
while from Eq. \eqref{eq3b-1} one can obtain the multiple
measurement setting for the uncertainty relation of quantum
coherence
%%%%%%%%%%%%%%%%%%%%%%%%%%%
\begin{equation}\label{eq3d-4}
 \sum_{i=1}^N C_r(\mathcal{B}_i,\rho_{AB})\geq \log_2 \frac{1}{b} -S(A|B),
\end{equation}
%%%%%%%%%%%%%%%%%%%%%%%%%%%
where the reference basis $\mathcal{B}_i= \{|\psi_k^{M_i}\rangle
\otimes |\varphi_l^B \rangle\}$.

\subsection{Information exclusion relation}
Formulated initially by Hall \cite{ier1}, the information exclusion
relation is also an important notion in information theory, and can
be obtained in a similar line to obtain the EURs. It quantifies the
amount of accessible information about the ensemble $\mathcal
{E}=\{p_i,\rho_i\}$ for observables $Q$ and $R$, and can be
described by the following inequality \cite{ier1}
%%%%%%%%%%%%%%%%%%%%%%%%%%%
\begin{equation}\label{eq3e-1}
 I(Q|\mathcal {E})+I(R|\mathcal {E}) \leq \log_2( d^2 c).
\end{equation}
%%%%%%%%%%%%%%%%%%%%%%%%%%%
where
%%%%%%%%%%%%%%%%%%%%%%%%%%%
\begin{equation}\label{eq3e-2}
 I(Q|\mathcal {E})= H(Q)_{\rho}-\sum_i p_i H(Q)_{\rho_i},
\end{equation}
%%%%%%%%%%%%%%%%%%%%%%%%%%%
is the Holevo quantity.

Subsequently, the information exclusion relation has been studied by
several other authors \cite{ier2,ier3}. In particular, Grudka
\emph{et al.} \cite{ier4} conjectured the following relation
%%%%%%%%%%%%%%%%%%%%%%%%%%%
\begin{equation}\label{eq3e-3}
 I(Q|\mathcal {E})+I(R|\mathcal {E}) \leq \log_2 \Bigg(d \sum_{d~\text{largest}}c_{ij}\Bigg).
\end{equation}
%%%%%%%%%%%%%%%%%%%%%%%%%%%
with the summation being taken over the largest $d$ terms of
$\{c_{ij}\}$. This bound is obviously tighter than that of Eq.
\eqref{eq3e-1} as $\sum_{d~\text{largest}}c_{ij}\leq d c$.

Coles and Piani \cite{bound1} further generalized the information
exclusion relation by replacing the ensemble $\mathcal{E}$ with a
quantum memory $B$, and proved that
%%%%%%%%%%%%%%%%%%%%%%%%%%%
\begin{equation}\label{eq3e-4}
 I(Q: B)+I(R: B) \leq \log_2 (d^2 c)- S(A|B),
\end{equation}
%%%%%%%%%%%%%%%%%%%%%%%%%%%
and
%%%%%%%%%%%%%%%%%%%%%%%%%%%
\begin{equation}\label{eq3e-5}
 I(Q: B)+I(R: B) \leq  r- S(A|B),
\end{equation}
%%%%%%%%%%%%%%%%%%%%%%%%%%%
where $r=\min\{r(Q,R),r(R,Q)\}$, and
%%%%%%%%%%%%%%%%%%%%
\begin{equation}\label{eq3e-6}
 \begin{split}
  & r(Q,R)= \log_2 \left(d\sum_j\max_k c_{jk}\right),\\
  & r(R,Q)= \log_2 \left(d\sum_k\max_j c_{jk}\right).
 \end{split}
\end{equation}
%%%%%%%%%%%%%%%%%%%%
The above bound is tighter than those of Eqs. \eqref{eq3e-1} and
\eqref{eq3e-3}. It can also be used to prove that Grudka \emph{et
al.}'s conjecture is right \cite{bound1}.

For the multiple measurement setting, Yu \emph{et al.} \cite{Yucs}
further generalized the information exclusion relation to
%%%%%%%%%%%%%%%%%%%%%%%%%%%
\begin{equation}\label{eq3e-7}
 \sum_{i=1}^N I(M_i:B) \leq \sum_{i=1}^N H(M_i)-\mathcal{L}_1,
\end{equation}
%%%%%%%%%%%%%%%%%%%%%%%%%%%
where $\mathcal{L}_1$ denotes the term on the RHS of Eq.
\eqref{eq3b-4}.

%\section{The uncertainty in open systems and its manipulation}
%\subsection{Effect of Noises on the entropic uncertainty}
%\subsubsection{Unital and nonunital noises}
%\subsubsection{Markovian and non-Markovian noises}
%\subsection{controlling of the entropic uncertainty}
%\subsection{Uncollapsing operation approach}
%\subsubsection{Weak measurement}
%\subsubsection{Filtering operation}
%\subsection{Non-Hermite operation approach}%
%\newpage
%\section{Applications}
%\subsection{Entanglement witness}
%\subsection{quantum  cryptography}
%\subsection{quantum speedup}
%\subsection{quantum  metrology}
%\subsection{quantum randomness}
%\subsection{ Quantum key distribution}
%\subsection{ others}

\section{The dynamical uncertainty in open systems and its manipulation} \label{sec:4}

\subsection{Effect of noises on the entropic uncertainty}

From a practical point of view, quantum objects are not isolated from others in general, hence are considerably fragile due to the decoherence of the system. In principle, the decoherence ought to affect the amount of the uncertainty more or less. In this sense, getting some insights into how the environment affects the uncertainty's magnitude becomes indispensable and crucial during quantum measurement. Up to now, much effort has been paid to unveil the quantum-memory-assisted EURs under the various environmental noises.

\subsubsection{Unital and nonunital noises}

As for the effects of noises on the entropic uncertainty, Xu \emph{et al}. \cite{Z.Y.Xu} examined dynamics of the entropic uncertainty in the unital and nonunital noisy channels, respectively. When one system to be probed is subjected to the noisy channel $\Lambda$, the state was mapped into
$\Lambda (\rho_0 )= \sum\nolimits_i {{E_i}\rho_0 E_i^\dag }$, with $E_i$ being the Kraus operator.
For the Bell-diagonal state
%%%%%%%%%%%%%%%%%%%%%%%%%%%
\begin{equation}
{\rho_{{\rm{Bell}}}}=\frac{1}{4}\left({{\mathds{1}_A} \otimes {\mathds{1}_B}+ \sum\limits_{j=1}^3 {{C_{{\sigma_j}}}\sigma_j^A \otimes \sigma _j^B} } \right),
\end{equation}
%%%%%%%%%%%%%%%%%%%%%%%%%%%
if one measures the Pauli observable $\sigma _j$ and $\sigma _k$ $(j\neq k)$, the complementarity $c=1/2$, therefore the entropic uncertainty can be expressed as
%%%%%%%%%%%%%%%%%%%%%%%%%%%
\begin{align}
U{\rm{ = }}{H_{{\rm {bin}}}}\left( {\frac{{{C_{{\sigma _j}}}+1 }}{2}} \right) + {H_{{\rm{bin}}}}\left( {\frac{{{C_{{\sigma _k}}} +1}}{2}} \right).
\label{Eq.1}
\end{align}
%%%%%%%%%%%%%%%%%%%%%%%%%%%
When the initial Bell-diagonal state meets ${C_{{\sigma _j}}} =  - {C_{{\sigma _i}}}{C_{{\sigma _k}}}$ $ ( {i \ne j \ne k} )$, which is called the state preparation and measurement choice (SPMC), it is found that the lower bound ($U_L$) can  coincide with the entropic uncertainty ($U_L$), {\it i.e.}, $U_L=U_R$. It means that one can directly employ the system's joint entropy $S\left( {A|B} \right)$ to measure the degree of entropic uncertainty. The local unital noisy channels satisfy the unital condition
%%%%%%%%%%%%%%%%%%%%%%%%%%%
\begin{align}
\Lambda _n^A\left( {\frac{1}{d}{\mathds{1}_A}} \right) = \frac{1}{d}{\mathds{1}_A},
\label{Eq.2}
\end{align}
%%%%%%%%%%%%%%%%%%%%%%%%%%%
with $d$ being the dimension of $\mathcal {H}_A$. The corresponding Kraus operators are $E_0^n = \sqrt {1 - p} \mathds{1}$ and $E_1^n = \sqrt p {\sigma _n}$, with $p$ being the occurrence probability of the noise and $n=1,2,3$ denoting bit-flip, bit-phase-flip, and phase-flip channels, respectively. Here, the bit-flip and phase-flip channels cannot break SPMC condition, which has been explored and revealed by Wang {\it et al}. \cite{D.Wang}. However, when the bit-phase-flip noise takes place, the SPMC condition will not be satisfied. Additionally, if the initially prepared bipartite state owns maximally mixed subsystems, such as Bell-diagonal state, the lower bound of entropic uncertainty will monotonously increase under local unital noise.

Huang \emph{et al}. \cite{A.J.Huang} investigated the nonunital and non-semiclassical local channels, such as the amplitude damping channel characterized by the Kraus operators $E_0^{AD} = {e^{ - \Gamma t/2}}\left| 0 \right\rangle \left\langle 0 \right| + \left| 1 \right\rangle \left\langle 1 \right|$ and $E_1^{AD} = \sqrt {1 - {e^{ - \Gamma t}}} $ $| 1 \rangle \langle 0 |$. After the particle $A$ passes through this channel, $\Lambda _i^A( {{\mathds{1}_A}}/2) = [ {{e^{ - \Gamma t}}| 0 \rangle \langle 0 | + }$ $ ( 2 - e^{ - \Gamma t} )| 1 \rangle \langle 1 |]/2$ does not meet the unital condition. Hence, the evolved state is not a Bell-diagonal state and the SPMC condition mentioned before is not satisfied, we generally have that $U_L\neq U_R$. Besides, the measured uncertainty may be decreased with time under the influence of the amplitude-damping channel. Ming \emph{et al.} \cite{Ming1} also examined the dynamics of the entropic uncertainty under local generalized amplitude damping noises with three realistic cases: one scenario is that the measured particle suffers from the noise while the particle serving as the quantum memory is free from any noises; another scenario is that the particle as quantum memory suffers from the noise while the measured particle does not; the last scenario is that both of the particles are affected by the noises. They found that the uncertainty shows analogous characters of the dynamical evolutions with respect to the three scenarios.
{{Besides, Karpat \cite{kar2} studied quantum-memory-assisted EUR with respect to two incompatible observables in correlated dephasing channels,
and Wang {\it et al}.  \cite{wgy1} also observed the dynamics of entropic uncertainty and its lower bound when the system is subject to
amplitude damping, phase-damping and depolaring channels, respectively.
Very recently, the circuit cavity quantum electrodynamics (QED) system affected by quantum noisy channels was considered to
examine the dynamics of quantum-memory-assisted entropic uncertainty relation \cite{jyh1}. }}

\subsubsection{Markovian and non-Markovian noises}
In general, we can consider the environment either in the Markovian or non-Markovian regime. If the information of a system flows from the system to the environment in one-way manner, we say the environment is  Markovian; Contrarily, if the information stored in the central system is bidirectionally flow between the system and the environment, then the environment is termed as a non-Markovian one.

 Wang {\it et al}.  \cite{D.Wang1} studied the dynamics of the entropic uncertainty without a quantum memory in the structured reservoir when a qubit undergoes a crossover of non-Markovian and Markovian regimes. The system is composed of a two-level atom coupled with a composite environment (one single-mode cavity and one hierarchical reservoir). The system Hamiltonian is depicted by
%%%%%%%%%%%%%%%%%%%%%%%%%%%
\begin{align}\label{Eq.3}
 {H_s} = {H_0} + {H_I},
\end{align}
%%%%%%%%%%%%%%%%%%%%%%%%%%%
where $H_0$ denotes the free Hamiltonian of the composite system and $H_I$ denotes the interaction Hamiltonian with respect to both the atom-cavity and the cavity-reservoir. The reduced dynamics for the atomic state can be written as
%%%%%%%%%%%%%%%%%%%%%%%%%%%
\begin{align}
\rho \left( t \right) = \left( {\begin{array}{*{20}{c}}
{{\rho _{ee}}\left( t \right)}&{{\rho _{eg}}\left( t \right)}\\
{\rho _{eg}^{\rm{*}}\left( t \right)}&{1 - {\rho _{ee}}\left( t \right)}
\end{array}} \right),
\label{Eq.4}
\end{align}
%%%%%%%%%%%%%%%%%%%%%%%%%%%
where${\rho _{ee}}\left( t \right){\rm{ = }}{\rho _{ee}}\left( 0 \right){\left| {\Gamma \left( t \right)} \right|^2}$ and ${\rho _{eg}}\left( t \right){\rm{ = }}{\rho _{eg}}\left( 0 \right)\Gamma \left( t \right)$  with $\Gamma ( t ){\rm{ = }}{L^{ - 1}} [ {\Upsilon ( p )} ]$, and ${L^{ - 1}}$ is the canonical inverse Laplace transformation. For a reservoir with memory effects, the uncertainty shows a quasi-periodic oscillation dynamic and reach the lower bound in the long-time limit with an arbitrary initial state $| {\psi } \rangle {\rm{ = }}\cos \theta | e  \rangle  + \sin \theta  {e^{i\phi }}  |g  \rangle $ mapping in the surface of the Block sphere. For a reservoir without memory effects, they claimed that the coupling strengths of the atom-cavity and the cavity-reservoir largely influence the magnitude of the uncertainty and its dynamical behaviors. The relatively strong coupling strength of the cavity and the structured reservoir can reduce the amount of uncertainty. That is, the relatively strong coupling strength between the atom and the cavity is responsible for the non-Markovianity. By contrast, the weak coupling strength will lead to the Markovianity. The stronger atom-cavity coupling strength leads to information backflow to the atom manifesting itself as an oscillation for the measured uncertainty. Notably, the uncertainty oscillates to the bound of the measured uncertainty when the coupling strength of the atom-cavity is stronger than the critical coupling strength $\Omega_{cr}$ \cite{D.Wang1}; the uncertainty will decrease all the time and reach the lower bound in the long-time limit when the atom-cavity coupling strength is weaker than the critical coupling strength.

Later, the EUR with quantum memory was discussed under the crossover between the non-Markovian and the Markovian regimes for a central system consists of two atoms independently coupled to the structured bosonic reservoirs \cite{D.Wang2}. Since a single "qubit+reservoir" Hamiltonian can be written as \cite{S.Maniscalco}
%%%%%%%%%%%%%%%%%%%%%%%%%%%
\begin{align}
H = {\omega _0}{\hat \sigma ^ + }{\hat \sigma ^ - } + \sum\limits_k {{\omega _k}\hat b_k^\dag \hat b_k + } \sum\limits_k {\left( {{g_k}{{\hat \sigma }^ + }{{\hat b}_k} + g_k^ * {{\hat \sigma }^ - }\hat b_k^\dag } \right)},
\label{Eq.5}
\end{align}
%%%%%%%%%%%%%%%%%%%%%%%%%%%
where ${\omega _0}$ denotes the qubit's transition frequency, ${\hat \sigma ^ + }\left( {{{\hat \sigma }^ - }} \right)$ is the raising (lowering) operator, ${\hat b_k}$ $ ( {\hat b_k^\dag } )$ is the creation (annihilation) operator, ${\omega _k}$ is the mode frequency of the {\it k}th field, and ${g _k}$ is the coupling strength. This model is solvable at the zero-temperature approximation and the qubit's dynamics can be characterized by the density matrix \cite{Huaop2343}
%%%%%%%%%%%%%%%%%%%%%%%%%%%
\begin{align}
{\hat \rho ^\alpha }\left( t \right) = \left( {\begin{array}{*{20}{c}}
{\rho _{ee}^\alpha \left( 0 \right){\zeta _t}}&{\rho _{eg}^\alpha \left( 0 \right)\sqrt {{\zeta _t}} }\\
{\rho _{ge}^\alpha \left( 0 \right)\sqrt {{\zeta _t}} }&{1 - \rho _{gg}^\alpha \left( 0 \right){\zeta _t}}
\end{array}} \right),
\label{Eq. 6}
\end{align}
%%%%%%%%%%%%%%%%%%%%%%%%%%%
for qubit $\alpha $, ${\zeta _t} =  - \int_0^t {d{t_1}f\left( {t - {t_1}} \right)} \zeta \left( t \right)$ with the correlation function $f\left( {t - {t_1}} \right)$. Considering the state
%%%%%%%%%%%%%%%%%%%%%%%%%%%
\begin{equation}\label{Eq. rhoab}
{\hat \rho _{AB}} = \frac{1}{4}\left( {{\mathds{1}_{AB}} + \vec{r} \cdot {\sigma ^A} \otimes {\mathds{1}_B} + {\mathds{1}_A} \otimes \vec{s} \cdot {\sigma ^A} + \sum\limits_{i = 1}^3 {{c_i}\sigma _i^A \otimes \sigma _i^B} } \right),
\end{equation}
%%%%%%%%%%%%%%%%%%%%%%%%%%%
with $\vec{r}= \vec{s} = 0$, the dynamics of the quantum-memory-assisted entropic uncertainty is considerably distinctive in the Markovian and non-Markovian regimes. The strong non-Markovianity can result in large-amplitude and long-period oscillations of the measured uncertainty and the lower bound. However, for the Markovian regime, the uncertainty and the lower bound will first increase and then subsequently reduce to a fixed value with time.
{{In addition, there exist several works \cite{M.N.Chen1,kar1,cpf222} to observe the dynamical characteristics of the entropy-based uncertainty affected by non-Markovianity.}}

\subsubsection{Dynamics of the EUR in specific systems}

\paragraph{The curved space-time}
$\\$

In 2013, Feng {\it et al.} \cite {J. Feng1} first observed the quantum-memory-assisted EURs in the frame of a Schwarzschild black hole. Typically, the Schwarzschild black hole is considered as offering one of the curved space time.
And it in Schwarzschild coordinates is described by
%%%%%%%%%%%%%%%%%%%%%%%%%%%
\begin{align}
d{s^2} =  - \left( {1 - \frac{{2M}}{r}} \right)d{t^2} + {\left( {1 - \frac{{2M}}{r}} \right)^{ - 1}}d{r^2} + {r^2}d{\Omega ^2},
\label{Eq. 7}
\end{align}
%%%%%%%%%%%%%%%%%%%%%%%%%%%
where $M$ denotes mass of the Schwarzschild black hole and $d{\Omega ^2}$ denotes the line element in the unit sphere. The observer outside the black hole cannot get the information regarding the particle's state inside the horizon in the presence of event horizon ${R_H} = 1/2M$. The information loss physically leads to a thermal feature for the vacuum in thermo-field dynamics and the vacuum is associated with  a spectrum with Hawking temperature. In Boulware basis, the Hartle-Hawking vacuum and its excitation are expressed as
%%%%%%%%%%%%%%%%%%%%%%%%%%%
\begin{align}
{\left| {{0_{{\omega _i}}}} \right\rangle _H} =& {\left[ {1 + \exp \left( { - \Omega \sqrt {1 - 1/{R_0}} } \right)} \right]^{ - \frac{1}{2}}}{\left| {{0_{{\omega _i}}}} \right\rangle _I}{\left| {{0_{{\omega _i}}}} \right\rangle _{II}} \nonumber \\
 &+ {\left[ {1 + \exp \left( { - \Omega \sqrt {1 - 1/{R_0}} } \right)} \right]^{ - \frac{1}{2}}}{\left| {{1_{{\omega _i}}}} \right\rangle _I}{\left| {{1_{{\omega _i}}}} \right\rangle _{II}},\nonumber \\
{\left| {{1_{{\omega _i}}}} \right\rangle _H} = &{\left| {{1_{{\omega _i}}}} \right\rangle _I}{\left| {{0_{{\omega _i}}}} \right\rangle _{II}},
\label{Eq. 8}
\end{align}
%%%%%%%%%%%%%%%%%%%%%%%%%%%
where  $R_0 = {r_0}/{R_H}$ with the position ${r_0}$  is in the vicinity of the event horizon, and $\Omega  = 2\pi \omega /\kappa {\rm{ = }}8\pi \omega M$ represents the measured mode frequency for the surface gravity. It can be found that the Hawking radiation can induce an important modification on the lower bound of the uncertainty. As to the uncertainty game between an observer freely falling and his/her static corporator possessing a quantum memory initially correlated to  quantum subsystem to be measured, thus information loss rooting from Hawking radiation inevitably lead to the increasing for the amount of uncertainty. The entropic uncertainty is sensitive to the mass of the black hole, the mode frequency of the quantum memory, and the distance of the observer from event horizon. Besides, to show the generality of their result, the entropic uncertainty is compared with other uncertainty measurement, {\it i.e.}, Aharonov-Anandan time-energy uncertainty.

%For a nonstationary quantum state $ | {\psi ( t  )}  \rangle $, the nonzero energy uncertainty reads
%%%%%%%%%%%%%%%%%%%%%%%%%%%%
%\begin{align}
%\Delta {E^2}\left( t \right) = \left\langle {\psi \left( t \right)} \right|{H^2}\left| {\psi \left( t \right)} \right\rangle  - {\left( {\left\langle {\psi \left( t \right)} \right|H\left| {\psi \left( t \right)} \right\rangle } \right)^2}.
%\label{Eq. 9}
%\end{align}
%%%%%%%%%%%%%%%%%%%%%%%%%%%

With regard to the uncertainty game between two static players, the measured  system $A$ holding by Alice and $B$  serving as quantum memory by Bob typically can be imitated by
 a pair of two-level atoms interacting with a bath of fluctuating massless quantum scalar fields outside the black hole. The systematic Hamiltonian is expressed by
%%%%%%%%%%%%%%%%%%%%%%%%%%%
\begin{align}
H{\rm{ = }}\frac{{{\omega _0}}}{2}\sum\limits_{i = 1}^3 {{n_i}{\Sigma _i}}  + {H_\phi } + {H_I},
\label{Eq. 10}
\end{align}
%%%%%%%%%%%%%%%%%%%%%%%%%%%
where ${\Sigma _i} = \sigma _i^A \otimes {\mathds{1}_B} + {\mathds{1}_A} \otimes \sigma _i^B$ denotes the symmetrized two-system operators, ${\omega _0}$ represents energy level spacing of the atoms, ${H_\phi }$ describes the Hamiltonian for the free massless scalar fields complying with Klein-Gordon equation in the outer of a black hole, and $H{}_I$ represents the interaction between the atoms and the bath.
The composite system finally will reach an equilibrium. As a matter of fact, the quantum information of subsystem {\it A} is transferred and stored in the quantum memory by the produced entanglement between them.
Notably, the entanglement can be witnessed via  $S\left( {A|B} \right)<0$.

Lately, Huang {\it et al.} \cite{J.L.Huang} studied the EUR towards the Dirac fields with and without spin, in the vicinity of the event horizon of a Schwarzschild black hole and proved that the bounds can be rewritten by means of the Holevo quantity as
%%%%%%%%%%%%%%%%%%%%%%%%%%%
\begin{align}
H\!\!\left( {{M_1}|B} \right)\! \!+\!\! H\!\!\left( {{M_1}|B}\! \right)\! \ge\! \! - \!{\log _2}c \!\!+\!\! H\!\!\left( A \right)\! \!-\!\! \!J\!\!\left( {B|{M_1}} \right)\! \!-\!\! \!\!J\!\!\left( {B|{M_2}} \right),
\label{Eq. 11}
\end{align}
%%%%%%%%%%%%%%%%%%%%%%%%%%%
and the Holevo quantity $J\left( {B|{M_1}} \right) = H\left( B \right) - \sum\nolimits_j {p_j^1H ( {{\rho _{B|u_j^1}}}  )} $ bounds how much information encoded in a quantum system with the corresponding measurement $ |{{u_j}}\rangle $ and the probability ${p_j} = \mathrm{Tr}({\langle {{u_j}} |\rho |{{u_j}} \rangle})$. The results reflected that the Holevo bound is tighter than the previous bound versus mutual information. Additionally, when the quantum memory goes away from the black hole, the discrepancy between the uncertainty and the proposed lower bound becomes invariable, and will not depend on any properties of the black hole.

Moreover, the quantum-memory-assisted EUR for the Dirac particles in the background of a Garfinkle-Horowitz-Strominger (GHS) dilation black hole have been studied {{\cite{Z.Y.Zhang1,ming22}}}. Generally, the spherically symmetric line element of another black hole (GHS dilation space-time) can be given by
%%%%%%%%%%%%%%%%%%%%%%%%%%%
\begin{align}
d{s^2} = &r\left( {r - 2D} \right)\left( {d{\theta ^2} + {{\sin }^2}\theta d{\phi ^2}} \right) - \left( {\frac{{r - 2M}}{{r - 2D}}} \right)d{t^2} \nonumber \\
&+ {\left( {\frac{{r - 2M}}{{r - 2D}}} \right)^{ - 1}}dr{}^2,
\label{Eq. 12}
\end{align}
%%%%%%%%%%%%%%%%%%%%%%%%%%%
where $M$ and $D$  denote parameters with respect to the mass of black hole and dilation field, respectively. After properly normalizing the state vector, we can obtain the vacuum state and excited state of the Kruskal particle for mode $\vec{k}$
\begin{equation}
\begin{aligned}\left|0_{\mathbf{k}}\right\rangle_{k}^{+}=&\left[1+e^{-8(M-D) \pi w_{i}}\right]^{-\frac{1}{2}}\left|0_{\mathbf{k}}\right\rangle_{I}^{+}\left|0_{\mathbf{k}}\right\rangle_{I I}^{-} \\ &+\left[1+e^{8(M-D) \pi w_{i}}\right]^{-\frac{1}{2}}\left|1_{\mathbf{k}}\right\rangle_{I}^{+}\left|1_{\mathbf{k}}\right\rangle_{I I}^{-} \\\left|1_{\mathbf{k}}\right\rangle_{k}^{+}=&\left|1_{\mathbf{k}}\right\rangle_{I}^{+}\left|0_{-\mathbf{k}}\right\rangle_{I I}^{-} \end{aligned}
\label{Eq. 13}
\end{equation}
%\begin{align}
%| {0_\vec{k}}\rangle _k^ +  =& {[ {1 + {e^{ - 8(M  -  D)\pi {w_i}}}}]^{-\frac12}}\left| {{0_\vec{k}}} \right\rangle _I^ + | {{0_\vec{k}}} \rangle _{II}^ - \nonumber \\
%&+  {\left[ {1  +   {e^{8\left( {M  -  D} \right)\pi {w_i}}}} \right]^{-\frac12}}\left| {{1_\vec{k}}} \right\rangle _I^ + \left| {{1_\vec{k}}} \right\rangle _{II}^ - ,\nonumber \\
%\left| {{1_\vec{k}}} \right\rangle _k^ +  = &\left| {{1_\vec{k}}} \right\rangle _I^ + \left| {{0_{ -\vec{k} }}} \right\rangle _{II}^ -,
%\label{Eq. 13}
%\end{align}
where ${\omega _i}$  is frequency, $ \{ { | {{m_{ \pm \vec{k} }}} \rangle _{I,II}^ \pm }  \}$ corresponds to the orthonormal bases for the outside and inside regions of the event horizon, respectively. The superscripts $\left\{  \pm  \right\}$  indicates the particle and antiparticle vacuum. Supposing that a hybrid qubit-qutrit initial state is prepared in \cite{Z.Y.Zhang1}
%%%%%%%%%%%%%%%%%%%%%%%%%%%
\begin{align}
\rho  = &\frac{{1 - 2p}}{2}\left( {\left| {01} \right\rangle \left\langle {01} \right| + \left| {01} \right\rangle \left\langle {20} \right|} \right.\left. { + \left| {20} \right\rangle \left\langle {01} \right| + \left| {20} \right\rangle \left\langle {20} \right|} \right)\nonumber \\
&{\rm{   }} + \frac{p}{2}\left( {\left| {00} \right\rangle \left\langle {00} \right| + \left| {00} \right\rangle \left\langle {21} \right| + \left| {10} \right\rangle \left\langle {10} \right| + \left| {11} \right\rangle \left\langle {11} \right|} \right.\nonumber \\
&\left. {{\rm{   }} + \left| {21} \right\rangle \left\langle {00} \right| + \left| {21} \right\rangle \left\langle {21} \right|} \right).
\label{Eq. 14}
\end{align}
%%%%%%%%%%%%%%%%%%%%%%%%%%

The entanglement decreases monotonously with the increase of the state parameter $p$  which range form 0 to 1/3. It is considered that the quantum memory locates near the event horizon of a GHS-dilation black hole as a qubit and the measured particle stays at the asymptotically flat region as a qutrit. It can be obtained that the uncertainty in the physically accessible region enlarges with the increasing dilation parameter of the black hole, whereas the  uncertainty in the inaccessible region reduces. Besides, to reveal the relationship between the entropic uncertainty and the system entanglement, the negativity is employed as the characterization of the distillable entanglement between the measured particle and the quantum memory. The negativity can be formulated from Peres criterion of separability as \cite{vg}
%%%%%%%%%%%%%%%%%%%%%%%%%
\begin{equation}
{\cal  N}\left( \rho  \right) = \sum\limits_i {\left| {{\lambda _i}\left( {{\rho ^{{T_A}}}} \right)} \right| - 1},
\end{equation}
%%%%%%%%%%%%%%%%%%%%%%%%%
where ${\lambda _i} ( {{\rho ^{{T_A}}}} )$ represents the \emph{i}th eigenvalue of the partial transpose matrix ${\rho ^{{T_A}}}$. It shows that the dynamical behavior of uncertainty is anti-correlated with the system's entanglement.

$\\$
\paragraph{The noninertial frame}
$\\$

The fermionic modes under Unruh effect can be depicted by the Rindler coordinates. The fermionic modes is divided into two Rindler wedges through acceleration horizon. Since the different wedges of field modes is restricted and uncorrelated, the information loss for the accelerated observer results in a thermal bath. The Unruh vacuum state ${\left| {{0_\omega }} \right\rangle _U}$ and one-particle state ${\left| {{1_\omega }} \right\rangle _U}$ can be written explicitly as
%%%%%%%%%%%%%%%%%%%%%%%%
\begin{equation}
\begin{split}
{\left| {{0_\omega }} \right\rangle _U} =& {\cos ^2}\alpha \left| {{0_\omega }} \right\rangle _I^ + \left| {{0_\omega }} \right\rangle _I^ - \left| {{0_\omega }} \right\rangle _{II}^ + \left| {{0_\omega }} \right\rangle _{II}^ -  \\
& - \cos \alpha \sin \alpha \left| {{0_\omega }} \right\rangle _I^ + \left| {{1_\omega }} \right\rangle _I^ - \left| {{0_\omega }} \right\rangle _{II}^ + \left| {{1_\omega }} \right\rangle _{II}^ - \\
& + \cos \alpha \sin \alpha \left| {{1_\omega }} \right\rangle _I^ + \left| {{0_\omega }} \right\rangle _I^ - \left| {{1_\omega }} \right\rangle _{II}^ + \left| {{0_\omega }} \right\rangle _{II}^ -   \\
& - {\sin ^2}\alpha \left| {{1_\omega }} \right\rangle _I^ + \left| {{1_\omega }} \right\rangle _I^ - \left| {{1_\omega }} \right\rangle _{II}^ + \left| {{1_\omega }} \right\rangle _{II}^ -,   \\
\left| {{1_\omega }} \right\rangle _U^ +  = &{q_R}\cos \alpha \left| {{1_\omega }} \right\rangle _I^ + \left| {{0_\omega }} \right\rangle _I^ - \left| {{0_\omega }} \right\rangle _{II}^ + \left| {{0_\omega }} \right\rangle _{II}^ -   \\
& - {q_R}\sin \alpha \left| {{1_\omega }} \right\rangle _I^ + \left| {{1_\omega }} \right\rangle _I^ - \left| {{0_\omega }} \right\rangle _{II}^ + \left| {{1_\omega }} \right\rangle _{II}^ -   \\
& + {q_L}\sin \alpha \left| {{1_\omega }} \right\rangle _I^ + \left| {{0_\omega }} \right\rangle _I^ - \left| {{1_\omega }} \right\rangle _{II}^ + \left| {{1_\omega }} \right\rangle _{II}^ -   \\
& - {q_L}\cos \alpha \left| {{0_\omega }} \right\rangle _I^ + \left| {{0_\omega }} \right\rangle _I^ - \left| {{0_\omega }} \right\rangle _{II}^ + \left| {{1_\omega }} \right\rangle _{II}^ -
\label{Eq. 15}
\end{split}
\end{equation}
%%%%%%%%%%%%%%%%%%%%%%%%
for the fermionic case. The superscripts $ \pm $ denote particle and anti-particle, and the subscripts {\it I} and {\it II} are the Rinder regions {\it I} and {\it II}, respectively. $q_R$ and $q_L$ are complex values and satisfy the normalized condition ${ | {{q_R}} |^2} + { | {{q_L}}  |^2} = 1$. The dimensionless acceleration parameter $\alpha $  is described as $\tan \alpha  = \exp \left( { - \pi \omega /a} \right)$, with $a \in \left[ {0} \right.\left. \infty  \right)$ and $\omega $  represents the frequency of the Unruh mode.

To explore the collective influence of the Unruh effect and the generalized amplitude damping noise or the phase-bit-flipping noise on the entropic uncertainty, the composite system of Alice and Bob has been considered with a generic Werner state \cite{D.Wang3}. It exposes that Unruh effect from the acceleration of assisted quantum memory can decrease the quantum correlation of bipartite system in the physical accessible region {\it I}, and consequently increase the amount of uncertainty. The explorations reveal that the system's information is redistributed, and some of the total available information flows towards the physically inaccessible region {\it II}. Note that the uncertainty  saturates into a constant in the limit of infinite acceleration $a$. Furthermore, the influence of Unruh effect from the acceleration on the uncertainty is larger than the noises. It is also gained that the unital noises can decrease the uncertainty in long-time regime.

$\\$
\paragraph{The single nitrogen-vacancy center in diamond}
$\\$

The single nitrogen-vacancy (NV) center in diamond is composed of the electron spin and the nuclear spin. Xu {\it et al.} \cite{APL} proposed a scheme to test the quantum-memory-assisted EUR in a single NV center in diamond only by performing local electronic measurements. The electron spin is treated as the measured object while the nuclear spin is treated as the quantum memory. As an application, the EUR is employed to witness entanglement between the electron  spin and the nuclear spin of the NV center. Remarkably, they displayed a specific numerical solution for the entropic uncertainty and the bound of entropic uncertainty for an arbitrary two-qubit initial state, which can be written as
%%%%%%%%%%%%%%%%%%%%%%%%%%
\begin{equation}
\begin{split}
{\rho _{AB}} = &\frac{1}{4}\left[ {{\mathds{1}_A} \otimes {\mathds{1}_B}} + \sum _{i = 1}^3 {\left( {{a_i}\sigma _i^A \otimes {\mathds{1}_B} + {\mathds{1}_A} \otimes {b_i}\sigma _i^B} \right)}\right.   \\
&\left.{ + \sum _{i,j = 1}^3 {{T_{ij}}\sigma _i^A \otimes \sigma _j^B} } \right],
\label{Eq. 16}
\end{split}
\end{equation}
%%%%%%%%%%%%%%%%%%%%%%%%%%
where ${\sigma _{i ( j  )}}$ ($i, j \in \left\{ {1,2,3} \right\}$) represent the Pauli operators, and the vectors {\it{a}} and {\it{b}} are defined with real components ${a_i}  = {\mathrm{Tr} }  (  {{\rho _{AB}} \sigma_i^A  \otimes {\mathds{1}_B}} )$ and ${b_i}  = {\mathrm{Tr} } (  {{\rho _{AB}}  {\mathds{1}_A}  \otimes \sigma_i^B} )$, respectively. The correlation tensor ${\it{T}}$  is defined by real components ${T_{ij}} = {\mathrm{Tr} } ( {{\rho _{AB}}\sigma _i^A \otimes \sigma _j^B} )$. If we measure two of the Pauli observables on particle {\it A}, the entropic uncertainty can be analytically obtained as
%%%%%%%%%%%%%%%%%%%%%%%%%%
\begin{align}
{U_L} = \sum\limits_{\scriptstyle x,y = 0,1\hfill\atop
\scriptstyle\lambda  = 1,3\hfill} {\eta _{xy}^\lambda {{\log }_2}\eta _{xy}^\lambda  - 2{H_\mathrm{bin}}\left( {\frac{{1 - \left\| b \right\|}}{2}} \right)},
\label{Eq. 17}
\end{align}
%%%%%%%%%%%%%%%%%%%%%%%%%
where
%%%%%%%%%%%%%%%%%%%%%%%%%%
\begin{equation}
\eta _{xy}^\lambda {\rm{ = }}\frac14 \left[ {1 + {{\left( { - 1} \right)}^x}{a_\lambda }}  + {{{\left( { - 1} \right)}^y}\sqrt {\sum\nolimits_{i = 1}^3 {{{\left({{y_i} + {{\left({ - 1} \right)}^x}{T_{\lambda i}}} \right)}^2}} }} \right],
\end{equation}
%%%%%%%%%%%%%%%%%%%%%%%%%%
and $\left\| b \right\|{\rm{=}} \left[{\sum\nolimits_{i = 1}^3 {b_i^2}}\right]^{1/2} $. Because the complementarity $c$  for the Pauli observables is 1/2, the lower bound of the uncertainty  is given by
%%%%%%%%%%%%%%%%%%%%%%%%%%
\begin{align}
{U_R} = S\left( {{\rho _{AB}}} \right) - {H_\mathrm{bin}}\left( {\frac{{1 - \left\| b \right\|}}{2}} \right) + 1.
\label{Eq. 18}
\end{align}
%%%%%%%%%%%%%%%%%%%%%%%%%%

$\\$
\paragraph{The spin-chain systems}
$\\$

Here, we mainly concentrate on the EUR in various spin-chain systems. For the chain with nearest-neighbor interactions, the Hamiltonian for an one-dimensional Heisenberg \emph{XYZ} chain can be expressed as
%%%%%%%%%%%%%%%%%%%%%%%%%%
\begin{align}
{H} = \frac{1}{2}\sum\limits_{k = 1}^n {\left( {{J_x}\sigma _k^x\sigma _{k + 1}^x + {J_y}\sigma _k^y\sigma _{k + 1}^y + {J_z}\sigma _k^z\sigma _{k + 1}^z} \right)},
\label{Eq. 18}
\end{align}
%%%%%%%%%%%%%%%%%%%%%%%%%%
where $\sigma _k^\gamma$ $  ( {\gamma {\rm{ = }}x,y,z} )$ is the Pauli operator at site $k$ and ${J_\gamma }$ is a real coupling strength with respect to the spin-spin interaction. If ${J_x} = {J_y}$ and ${J_z} = 0$, the corresponding Heisenberg chains is called the {\it XX} model, and its system Hamiltonian for a two-spin system was given by
%%%%%%%%%%%%%%%%%%%%%%%%%%
\begin{align}
{{ H}_{AB}} = \frac{1}{2}\left[ {{J_x}\sigma _1^x\sigma _2^x + {J_y}\sigma _1^y\sigma _2^y +  ( {G + g} )\sigma _1^z +  ( {G -g} )\sigma _2^z} \right],
\label{Eq. 19}
\end{align}
%%%%%%%%%%%%%%%%%%%%%%%%%%
when an inhomogeneous magnetic field is applied along the {\it z}-direction, where $G$ and $g$ correspond the degree of inhomogeneity and the degree of inhomogeneity, respectively. The quantum-memory-assisted EUR was first studied in a two-qubit Heisenberg $XX$ model by Huang {\it et al.} \cite{A.J.Huang3}. Their results reflect that the larger coupling coefficient between two spin qubits can decrease the uncertainty of interest and the entropic uncertainty will even reach to zero for the relatively large coupling coefficients. Moreover, the entropic uncertainty presents various characteristic when $g< 1$ and $g > 1$. The relation between the entropic uncertainty, the purity $P = \mathrm{Tr} \left({\rho _{AB}^2}\right)$ and the Bell non-locality are compared. It was found that the entropic uncertainty is anti-correlated with both the purity and the Bell non-locality. Afterwards, there are some studies on the quantum-memory-assisted EUR in other Heisenberg spin-chain models and Heisenberg models with Dzyaloshinski-Moriya (DM) interaction. {For example}, the dynamical characteristics of the quantum-memory-assisted EUR have been observed in the canonical Heisenberg {\it XXX}\cite{D.Wang5} and {\it XXZ}\cite{Ming2} models with an inhomogeneous magnetic field.
{Afterwards, Wang {\it et al.} \cite{D.Wang4} explored the relation between the entropic uncertainty and quantum correlation in the general Heisenberg {\it XYZ} model with an inhomogeneous magnetic field. It is   worth noting that an interesting result can be obtained
\begin{align}
{U_R} &= S\left( {{\rho _{AB}}} \right) - S\left( {{\rho _B}} \right) + {\log _2}\frac{1}{c} \nonumber \\
&= {\log _2}\frac{1}{c}{\rm{ + }}\mathop {{\rm{min}}}\limits_{\{ {\Pi _{_i}^B} \}} {\rm{[}}{S_{\{ {\Pi _{_i}^B} \}}}\left( {{\rho _{A|B}}} \right)] - D\left( {{\rho _{AB}}} \right),
\label{Eq. 200}
\end{align}
which clearly shows that the entropic uncertainty's lower bound $U_R$ in  Eq. \eqref{eq3-1} is anti-correlated to the term, {\it i.e.}, quantum correlation
$D\left( {{\rho _{AB}}} \right)$. When two observables are set, the correlation is not the only decisive factor of the bound, and the minimal von Neumann conditional entropy
$\mathop {{\rm{min}}}\limits_{\{ {\Pi _{_i}^B} \}} {\rm{[}}{S_{\{ {\Pi _{_i}^B} \}}}\left( {{\rho _{A|B}}} \right)]$ is another factor to determine the bound.}

In addition, Zheng {\it et al.} \cite{X.Zheng}and {{Huang {\it et al.} \cite{hzm11}}} investigated the relation between the system's entanglement and the lower bound, tightness of the entropic uncertainty in the Heisenberg model with DM interaction. Ming {\it et al.} \cite{Ming3} compared the effect of different components of the DM interaction on reducing the entropic uncertainty.
{{And it is found that the lower bound of entropic uncertainty $U_R$ in  Eq. \eqref{eq3-1} can be rewritten as
\begin{align}
{U_R} &= S\left( {{\rho _{AB}}} \right) - S\left( {{\rho _B}} \right) +  + {\log _2}\frac{1}{c} \nonumber \\
&=  - {C_r}\left( {{\rho _{AB}}} \right) + S\left( {{\rho _d}} \right) - S\left( {{\rho _B}} \right) + {\log _2}\frac{1}{c},
\label{Eq. 201}
\end{align}
where ${\rho _d} = \sum\limits_i {{\rho _{ii}}\left| i \right\rangle \left\langle i \right|}$ is the diagonal part of $\rho$. This shows that the uncertainty bound is closely anti-correlated with the quantum coherence, but not fully dependent on quantum coherence.}}
Yang {\it et al.} \cite{Y.Y.Yang} also studied dynamical characteristics of the entropic uncertainty in a general Heisenberg {\it XYZ} model with DM interaction. More recently,
 Zhang {\it et al.} \cite{zhangz22} and Shi {\it et al.} \cite{swn11} investigated quantum-memory-assisted EUR in the higher-dimensional Heisenberg model.

\subsection{Controlling the entropic uncertainty}
In reality, any quantum system will unavoidably interacts with the surroundings, resulting in decoherence or dissipation. With this in mind, how to effectively suppress decoherence is in demand when performing quantum tasks. To get the precise outcome for a measurement, it is expected that the measured uncertainty could be minimal in realistic quantum information processing. Motivated by this consideration, some researchers have devoted to pursuing various working strategy to manipulate the magnitude of the uncertainty.

\subsubsection{Uncollapsed measurement}
Two non-unitary quantum operations ({\it i.e.}, the quantum weak measurement and weak reversal measurement) are proposed to reduce the entropic uncertainty, {\it i.e.}, a class of uncollapsing operations {{\cite{Y.L.Zhang,cpf22,hs11}}}. The quantum weak measurement can be mathematically described by the following operators
%%%%%%%%%%%%%%%%%%%%%%%%%
\begin{align}
{M_w} = \left( {\begin{array}{*{20}{c}}
{\sqrt {1 - {k_w}} }&0\\
0&1
\end{array}}\right),\
{M_r} = \left( {\begin{array}{*{20}{c}}
1&0\\
0&{\sqrt {1 - {k_r}} }
\end{array}} \right),
\label{Eq. 7}
\end{align}
%%%%%%%%%%%%%%%%%%%%%%%%
where ${k_w} \in \left( {0,1} \right)$ and ${k_r} \in \left( {0,1} \right)$  are the weak measurement strength and weak measurement reversal strength, respectively. The final state can be written as
%%%%%%%%%%%%%%%%%%%%%%%%%%%%%
\begin{equation}
{\rho _{AB}}(t) = \frac{ { \left({M^A} \otimes {M^B}\right) {\rho _{AB} }{{\left({M^A} \otimes   {M^B}\right)}^\dag } } }{ \mathrm{Tr}  \left\{ {({M^A} \otimes {M^B}) {\rho  _{AB} }{{({M^A} \otimes {M^B})}^\dag } }\right \}}.
\end{equation}
%%%%%%%%%%%%%%%%%%%%%%%%%%%%%
Here, for a pure state $ | \psi   \rangle  = \cos \alpha {\left| {00} \right\rangle _{AB}} + \sin \alpha {\left| {11} \right\rangle _{AB}}$ with $\alpha\in \left[ {0,\pi /2} \right]$, two schemes have been proposed to govern the uncertainty in the noisy environment by utilizing prior weak measurement
and posterior weak measurement reversal.
%Scheme 1: only prior weak measurements are performed before the system {\it A} and
%system memory {\it B} being put into the dissipative environment.
%Scheme 2: only posterior weak measurements are taken after qubits A and B
%locating in the Marlovian and non-Marlovian noisy environment.
It was obtained that the peak values of the entropic uncertainty can be effectively decreased by the prior weak measurement for long periods of time, whereas it is invalid for the wave minima of the entropic uncertainty. However, the posterior weak measurement can effectively reduce the wave minima values of the entropic uncertainty, but it is invalid for the peak of entropic uncertainty. This can be explained by the fact that the prior weak measurement can strengthen the robustness of the system against the decoherence by moving a qubit to its ground state, while a posterior weak measurement reversal will move a qubit to its excited state to strengthen the coherence of two-qubit system.

Recently, by considering the weak measurement and measurement reversal, Guo {\it et al.} \cite{Y.N.Guo} explored the degradation of the entropic uncertainty with respect to a high-dimensional state under the amplitude damping channel at finite temperature. The Kraus operators are
%%%%%%%%%%%%%%%%%%%%%%%%
\begin{equation}
\begin{split}
{M_W} = &\left( {\begin{array}{*{20}{c}}
1&0&0\\
0&{\sqrt {1 - {k_W}} }&0\\
0&0&{\sqrt {1 - {k_W}} }
\end{array}} \right),  \\
{M_R} = &\left( {\begin{array}{*{20}{c}}
{1 - {k_R}}&0&0\\
0&{\sqrt {1 - {k_R}} }&0\\
0&0&{\sqrt {1 - {k_R}} }
\end{array}} \right).
\label{Eq. 8}
\end{split}
\end{equation}
%%%%%%%%%%%%%%%%%%%%%%%%
It has been verified that the uncertainty can be reduced for the posterior measurement reversal at the finite temperature, whereas the prior weak measurement can  decrease the uncertainty at relatively low temperatures but this becomes invalid at high channel temperatures.

\subsubsection{Filtering operation}
The filtering operation refers to the non-trace-preserving map and can be used to enhance entanglement of a system. In practice, this operation is usually deemed as one type of the weak measurement with null result \cite{Q.Su}. Explicitly, the filtering operation can be described by
%%%%%%%%%%%%%%%%%%%%%%%%%
\begin{align}
F = \left( {\begin{array}{*{20}{c}}
{\sqrt {1 - {k_f}} }&0\\
0&{\sqrt {{k_f}} }
\end{array}} \right),
\label{9}
\end{align}
%%%%%%%%%%%%%%%%%%%%%%%%
where ${k_f} \in [ {0,1} ]$ is the strength of the operation. Huang \emph{et al}. \cite{A.J.Huang} first studied the reduction of quantum-memory-assisted entropic uncertainty by performing the filtering operation, which is influenced by the unital and nonunital channels. It was shown that the entropic uncertainty will gradually decrease with the increase of the operation strength. It is because that the filtering operation is capable of suppressing the decoherence effect. There exists another vital explanation to illustrate the reduction of entropic uncertainty. The reason is that the filtering operation is a nonunitary operation, hence can exchange the information between the qubits {\it A} and {\it B} possibly. Thereby, the entropic uncertainty can be reduced probably.

\subsubsection{Non-Hermite operation}

It is usually required that the Hamiltonian of a physical system is Hermitian in conventional quantum mechanics, as this ensures that the system's energy is real and the time evolution is unitary. A class of Hamiltonian was proposed by Bender \emph{et al.} \cite{Bender} in 1998, {\it i.e.}, the so called parity-time $(\mathcal {P T})$ symmetric Hamiltonian with the parity reflection operator $\mathcal{P}$ and the time reversal operator $\mathcal {T}$. The $\mathcal {P T}$ symmetric Hamiltonian is non-Hermitian, whereas it still keeps the spectrum real. Later, they introduced a linear operator to construct a new inner product structure and this guaranteed that the time evolution operator is unitary. For a qubit, the $\mathcal{P T}$ symmetric Hamiltonian is given by
%%%%%%%%%%%%%%%%%%%%%%%%%
\begin{align}
{H_{\mathcal {P T}}} = s\left( {\begin{array}{*{20}{c}}
{i\sin \theta }&1\\
1&{ - i\sin \theta }
\end{array}} \right),
\label{10}
\end{align}
%%%%%%%%%%%%%%%%%%%%%%%%%
where $s$ and $\theta$ are real numbers, $s$  denotes a general scaling constant related to the matrix, $\theta \in [ { - \pi /2,\pi /2} ]$ represents the non-Hermiticity of the Hamiltonian. For $\theta  \to 0$, ${H_{\mathcal {P T}}}$ is Hermitian, and when $\theta  \to  \pm \pi /2$, ${H_{\mathcal {P T}}}$ is strongly non-Hermitian. The time-evolution operator of ${H_{\mathcal {P T}}}$  is given by
%%%%%%%%%%%%%%%%%%%%%%%%%
\begin{align} {U_{\mathcal {P T}}} = {e^{ - i{H_{\mathcal {P T}}}t}} = \frac{1}{{\cos \theta }}\left( {\begin{array}{*{20}{c}}
{\cos \left( {t' - \theta } \right)}&{ - i\sin t'}\\
{ - i\sin t'}&{\cos \left( {t' + \theta } \right)}
\end{array}} \right),
\label{10}
\end{align}
%%%%%%%%%%%%%%%%%%%%%%%%%
with $t' = \Delta {E_t}/2$, $\Delta E{\rm{ = }}{E_ + } + {E_ - }$, and ${E_ \pm } =  \pm s\cos \theta $ are the eigenvalues of $H_{\mathcal {P T}}$. Several experiments have demonstrated the ${\mathcal {P T}}$ symmetric Hamiltonian. Shi {\it et al}. \cite{W.N.Shi} explored the reduction of quantum-memory-assisted entropic uncertainty by means of the $\mathcal {P T}$ symmetric operation and gave a corresponding explanation.

%A two-qubit system is prepared in a maximally entangled state.
%After conventional local operation performing on qubit {\it A},
%the reduced density matrix of qubit {\it B} will remain in
%the maximally mixed state ${\rho _B} = {I_B}/2$. However,
%after implementing the local $\mathcal {P T}$ symmetric operation on
%qubit {\it A} with $\alpha  \ne 0$, the reduced density
%matrix will no longer remain in the maximally mixed state.
%It shows the property of quantum non-locality can
%make the information exchange between qubits {\it A}
%and {\it B} possibly. Hence, this can result in the degradation
%of the entropic uncertainty with a certain probability when the
%information flows form qubit {\it A} to {\it B}.

Besides, there are other approaches to control the entropic uncertainty, {\it e.g.}, Yu \emph{et al.} \cite{Yu1} proposed a strategy to control the entropic uncertainty in the presence of quantum memory via quantum-jump-based feedback, {{and Adabi \emph{et al.}\cite{adabi11} provided a method to degrade the lower bound  of the uncertainty via local operation and classical communication (LOCC)}}.

\section{Applications of the EURs} \label{sec:5}

\subsection{Entanglement witness}

Entanglement is a valuable physical resource in quantum information processing. Thereby, detecting entanglement is viewed as being a fundamental and indispensable task. Entanglement witness is the process which verifies that a source produces entangled states. A state is said to be entangle if it cannot be represented by a convex combination of product states. Practically, entanglement witness is to certify a mathematical identity condition and all separable states must satisfy identity condition. This identity condition is defined as an entanglement witness. If the source does not satisfy this identity condition, this means that the source can produce
entangled particles, which is demonstrated experimentally. Up to now, there are many types of entanglement witness operators being constructed \cite{G®πhne,Horodecki1}.

We hereafter mainly focus on entanglement witness via the EURs and these discussions are restricted to the bipartite entanglement. It is worth to state that the entanglement witness typically emerges in the paradigm of distant laboratories, in which a pair of observers can only make a local measurement on their own subsystem.

First of all, we introduce a simple and well-known entanglement witness with respect to the two-qubit system.
Incidentally, the introduced entanglement witness will rely on complementary observables other than entropy of interest. Hence we can  directly compare it with the entropic entanglement witnesses. Considering mutually unbiased observables, the operator is defined as
%%%%%%%%%%%%%%%%%%%%%%%%%%%%%%%%
\begin{align}
{E_{XZ}} = {E_X} + {E_Z},
\label{11}
\end{align}
%%%%%%%%%%%%%%%%%%%%%%%%%%%%%%%%
where ${E_X} = \left|  +  \right\rangle \left\langle  +  \right| \otimes \left|  - \right\rangle \left\langle  -  \right| + \left|  -  \right\rangle \left\langle  -  \right| \otimes \left|  +  \right\rangle \left\langle  +  \right|$ and ${E_Z} = \left| 1 \right\rangle \left\langle 1 \right| \otimes \left| 0 \right\rangle \left\langle 0 \right|+ \left| 0 \right\rangle \left\langle 0 \right| \otimes \left| 1 \right\rangle \left\langle 1 \right| $. When ${E_X} $ and ${E_Z}$  project onto the subspaces where two observers' measured outcomes are different, they are error operators. For the maximally entangled state
$ | \varphi \rangle {\rm{ = }}\left( {\left| {00} \right\rangle  + \left| {11} \right\rangle } \right)/\sqrt 2$, the error-probability is zero in either basis, that is, $ \langle \varphi  |{E_{XZ}} | \varphi  \rangle  = 0$. However, for any separable state ${\rho _{AB}}$, we have \cite{R.Namiki}
%%%%%%%%%%%%%%%%%%%%%%%%%%%%%%%%
\begin{align}
\mathrm{Tr}\left[ {{\rho _{AB}}{E_{XZ}}} \right] \geq \frac{1}{2}.
\label{12}
\end{align}
%%%%%%%%%%%%%%%%%%%%%%%%%%%%%%%%
Therefore, $\left\langle {{E_X}} \right\rangle  + \left\langle {{E_Z}} \right\rangle  < 1/2$ with $\left\langle \Xi  \right\rangle {\rm{ = }}\mathrm{Tr} \left[ {\Xi {\rho _{AB}}} \right]$ is defined as the linear witness criterion of entanglement.

Another criterion of bipartite entanglement witness is entropy based. For a quantum system consisting of two parties, we obtain that the system must be entangled if the conditional von Neumann entropy is negative, viz.
%%%%%%%%%%%%%%%%%%%%%%%%%%%%%%%%
\begin{align}
S(A|B)<0.
\label{5-13}
\end{align}
%%%%%%%%%%%%%%%%%%%%%%%%%%%%%%%%
This criterion can be connected to the lower bound of the EUR in Eq. (\ref{eq3-3}), and the Fano's inequality
%%%%%%%%%%%%%%%%%%%%%%%%%%%%%%%%
\begin{align}
S(Q|B)+S(R|B)<H_\mathrm{bin}(d_Q)+H_\mathrm{bin}(d_R),
\label{5-114}
\end{align}
%%%%%%%%%%%%%%%%%%%%%%%%%%%%%%%%
with the variable $d_Q$ denoting the probability that the outcomes of $Q$ on $A$ and $Q$ on $B$ are different \cite{Nphys1}. Thus, the quantum-memory-assisted EUR yields the criterion
%%%%%%%%%%%%%%%%%%%%%%%%%%%%%%%%
\begin{align}
H_\mathrm{bin}(d_Q)+H_\mathrm{bin}(d_R)<-\log_2 c.
\label{5-14}
\end{align}
%%%%%%%%%%%%%%%%%%%%%%%%%%%%%%%%
By taking the two mutually unbiased operators $Q$ and $R$ in a qubit system, one can attain that the system is entangled if the condition $H_\mathrm{bin}(d_Q)+ H_\mathrm{bin}(d_R)<1$ holds. Noteworthily, the criterion for entanglement witness is a sufficient and unnecessary condition, which implies that two particle may be in an entangled state when the inequality in Eq. (\ref{5-13}) is disobeyed.

\subsubsection{Shannon entropic witness}
Giovannetti \cite{VG} and G$\ddot{{\rm u}}$hne {\it et al.} \cite{OG} have done some early work about entanglement witness by means of the EURs, and Huang \cite{Hu112} made further improvements. We primarily review the entanglement witness on more recent developments from quantum-memory-assisted EURs. Berta {\it et al.} \cite{Berta} have discussed how to employ this relation to detect entanglement, while Li {\it et al.} \cite{Nphys1} and Prevedel {\it et al.} \cite{Nphys2} have verified this approach experimentally. To be explicit, it is shown that all separable states meet
%%%%%%%%%%%%%%%%%%%%%%%%%%%%%%%%
\begin{align}
S\left( {{X_A}|{X_B}} \right) + S\left( {{Z_A}|{Z_B}} \right) \ge {q_{MU}},
\label{13}
\end{align}
%%%%%%%%%%%%%%%%%%%%%%%%%%%%%%%%
where the quantity ${q_{MU}}$ is only related to Alice's observables. For separable states ${\rho _{AB}}$, the conditional entropy is nonnegative, {\it i.e.}, $S\left( {A|B} \right) \ge 0$. Based on the basis $\mathcal{X}_B$, the measured Bob's system can increase the magnitude of entropic uncertainty of Alice's measurement, that is $S\left( {{X_A}|{X_B}} \right) \ge S\left( {{X_A}|B} \right)$.
Suppose that Alice and Bob possess numerous
duplicates of the system state and both of them perform one measurement on each copy by one of two observables. From the joint probability distributions, it is easy to calculate the amounts of $S\left( {{X_A}|{X_B}} \right)$ and $S\left( {{Z_A}|{Z_B}} \right)$. If the inequality \eqref{13} is violated, it is sure that ${\rho _{AB}}$ is entangled. Comparing the efficiency of entanglement witness by Eq. (\ref{12}) and Eq. (\ref{13}), it can be found that the linear witness detects more entangled states than Shannon entropic witness. Whereas the quality of entanglement detected by Shannon entropy is higher because of the application of Eq. (\ref{13}) for all nondistillable states where Alice and Bob is unable to distill any EPR states by virtue of the local operation and classical communication (LOCC) \cite{Horodecki}. In other words, Shannon entropic witness can detect distillable entanglement, however, linear witness can detect all forms of entanglement. The coherent information $ - S\left( {A|B} \right)$ can bounds the distillable entanglement, which is employed to quantify the distillable entanglement  $E_D$ \cite{M.Berta2}
%%%%%%%%%%%%%%%%%%%%%%%%%%%%%%%%
\begin{align}
{E_D} \ge  - S\left( {A|B} \right).
\label{14}
\end{align}
%%%%%%%%%%%%%%%%%%%%%%%%%%%%%%%%
Combining the quantum-memory-assisted EUR, it is rewritten as
%%%%%%%%%%%%%%%%%%%%%%%%%%%%%%%%
\begin{align}
{E_D} \ge {q_{MU}} -S\left( {{X_A}|{X_B}} \right) +S\left( {{Z_A}|{Z_B}} \right).
\label{15}
\end{align}
%%%%%%%%%%%%%%%%%%%%%%%%%%%%%%%%
Notably, this result gives quantitative lower bound which displays a preponderance of Shannon entropic witness in contrast to the witness in Eq. (\ref{12}).

\subsubsection{Other entropic witnesses}

Berta {\it et al.} \cite{M.Berta2} also employed the uncertainty relation by the collision entropies to detect entanglement by adopting
 {\it k} MUBs on the system of Alice (a subset of size {\it k} of MUBs chosen from a set of ${d_A} + 1$ MUBs and  ${d_A}$ is a prime power and $2 \le k \le {d_A} + 1$). Consider such a set $\left\{ {{\mathcal{X}_j}} \right\}$ of $k$ MUBs on Alice's system, and consider a series of {\it k} arbitrary positive-operator-valued measurements $\left\{ {{\mathcal{Y}_j}} \right\}$ on Bob's system. It was shown that the condition of all separable states is
%%%%%%%%%%%%%%%%%%%%%%%%%%%%%%%%
\begin{align}
\sum\limits_{j = 1}^k {{2^{ - {S_2}\left( {{X_j}|{Y_j}} \right)}}}  \le 1 + \frac{{k - 1}}{{{d_A}}}.
\label{16}
\end{align}
%%%%%%%%%%%%%%%%%%%%%%%%%%%%%%%%
In contrast to the aforemention approaches, this entanglement witness can detect more entangled states than Shannon entropic witness, but not as much as the linear witness. Analogical to the Shannon entropic witness, it can get a quantitative lower bound of entanglement-like measurement from the uncertainty relation by the collision entropy. Later, Walborn {\it et al.} \cite {S.P.Walborn} extended the approach of entanglement witness by EURs to continuous variable systems, and Saboia {\it et al.} \cite {A.Saboia} and Huang {\it et al.} \cite {Y.Huang} made further explorations.

\subsection{Steering inequalities}
In 1935, Schr$\ddot{{\rm o}}$dinger first showed that steering is a phenomenon related to entanglement for bipartite systems, but it is not precisely the same to entanglement.
Taking into account the paradigm of distant laboratories where there are two participators Alice and Bob, owning
the subsystems  {\it A} and {\it B} respectively. Steering is denoted as that the measurement choice of one subsystem $A$ can result in
different ensembles of states on another subsystem {\it B}. Not all quantum states can show steering. For example, all separable states are nonsteerable.
In addition, Bell inequalities have been derived according to a local hidden variable (LHV) model. If a state violates a Bell inequality, it is steerable. In 2007, Wiseman {\it et al.} \cite {Wiseman1} formalized the notion of steerability for states getting rid of the LHV model, leading to a quantum state of subsystem $B$ being related to an
arbitrary observable of {\it A}. Based on this formalization, Cavalcanti {\it et al.} \cite {E.G.Cavalcanti} derived the steering inequalities in 2009.

Steering inequalities can be derived by using EURs. If subsystem {\it B} is dependent of a LHV, as a result,
the $B$'s measurement probabilities should submit to a single-system uncertainty relation on condition of the measured outcomes on subsystem {\it A}. To be precise, the local hidden state model yields the joint probability distribution for observables  ${\mathcal{X}_A}$ on {\it A} and  ${\mathcal{X}_B}$ on {\it B} with the form
%%%%%%%%%%%%%%%%%%%%%%%%%%%%%%%%
\begin{align}
p\left( {{\mathcal{X}_A},{\mathcal{X}_B}} \right) = \sum\limits_ \ell  {p\left( {\Omega  = \ell } \right)p\left( {{\mathcal{X}_A}|\Omega  = \ell  } \right){p_\kappa }\left( {{\mathcal{X}_B}|\Omega  = \ell  } \right)},
\label{17}
\end{align}
%%%%%%%%%%%%%%%%%%%%%%%%%%%%%%%%
where $\Omega $ corresponds to the hidden variable for Bob's local state, $\ell $ denotes a special value that the variable can reach, and
$\kappa $ is the probability distribution. Then we have
%%%%%%%%%%%%%%%%%%%%%%%%%%%%%%%%
\begin{equation}
\begin{split}
S\left( {{X_B}|{X_A}} \right) &\ge S\left( {{X_B}|{X_A}\Omega } \right)  \\
&= \sum\limits_\ell  {p\left( {\Omega  = \ell  } \right)S\left( {{X_B}|{X_A}\Omega  = \ell  } \right)}  \\
&= \sum\limits_\ell  {p\left( {\Omega  = \ell  } \right)S\left( {{X_B}|\Omega  =\ell  } \right)},
\label{18}
\end{split}
\end{equation}
%%%%%%%%%%%%%%%%%%%%%%%%%%%%%%%%
where $S\left( {{X_B}|{X_A}\Omega  = \ell } \right)$  represents the entropy of ${X_B}$ on condition of ${X_A}$ and the event $\Omega  = \ell  $. Hence, for two observables $\mathcal{X}_B$  and $\mathcal{Z}_B$ on {\it B}, and some other observables  $\mathcal{X}_A$  and $\mathcal{Z}_A$ on {\it A}, we have
%%%%%%%%%%%%%%%%%%%%%%%%%%%%%%%%
\begin{equation}
\begin{split}
S\left( {{X_B}|{X_A}} \right) + S\left( {{Z_B}|{Z_A}} \right) &\ge \sum\limits_\ell   {p\left( {\Omega  =\ell } \right)\left[ {S\left( {{X_B}|\Omega  = \ell  } \right)} \right.}  \\
 & + S( {{Z_B}|\Omega  = \ell })].
\label{19}
\end{split}
\end{equation}
%%%%%%%%%%%%%%%%%%%%%%%%%%%%%%%%
Combining the above inequation and Maassen-Uffink's uncertainty relation, one can derive the steering inequality by Schneeloch {\it et al.} \cite {J.Schneeloch}
%%%%%%%%%%%%%%%%%%%%%%%%%%%%%%%%
\begin{align}
S\left( {{X_B}|{X_A}} \right) + S\left( {{Z_B}|{Z_A}} \right) \ge {q_{MU}}.
\label{Eq.96}
\end{align}
%%%%%%%%%%%%%%%%%%%%%%%%%%%%%%%%
If a state admits a local hidden state model, then it must obey Eq. (\ref{Eq.96}).
Thereby, the violation of Eq. (\ref{Eq.96}) can be considered as a indicator of steering in the experiments.
Walborn {\it et al.} \cite {S.P.Walborn} also deduced similar steering inequalities for continuous variables.

\subsection{Wave-particle duality}

As is well know that wave-particle duality is a peculiar and remarkable characteristic of a single system. An interferometer cannot be designed to simultaneously show these two behaviors. Feynman qualitatively discussed this idea. Wootters {\it et al.}  \cite {Wootters},  Jaeger {\it et al.} \cite {Jaeger}, and Englert  \cite {Englert1} subsequently put on quantitative grounds and argued the inequalities
termed as wave-particle duality relations. Afterwards, a number of  the relevant relations were derived under the Mach-Zehnder interferometer for single photons.
In all of these cases,  the particle behaviors are related to the known paths that the photon passes  and the wave behaviors are related to seen oscillation in the probabilities to probe the photon in a specified output mode
when one changes the relative phase $\varphi$ between a pair of interferometer arms. $\mathcal{Z} = \left\{ {\left| 0 \right\rangle \left\langle 0 \right|,{\rm{ }}\left| 1 \right\rangle \left\langle 1 \right|} \right\}$  denotes the which-path observable, the path predictability ${\rm P} = 2{p_{{\rm{guess}}}}\left( Z \right) - 1$ can quantify particle behavior with the probability of precisely guessing the path ${p_{{\rm{guess}}}} ( Z  )$. The fringe visibility quantify the wave behavior
%%%%%%%%%%%%%%%%%%%%%%%%%%%%%%%%
\begin{align}
\Delta {\rm{ = }}\frac{{p_0^{\max } - p_0^{\min }}}{{p_0^{\max } + p_0^{\min }}},
\label{21}
\end{align}
%%%%%%%%%%%%%%%%%%%%%%%%%%%%%%%%
where $p_0^{\max } := \mathop {\max }_\Phi  {p_0}$ and $p_0^{\min } := \mathop {\min }_\Phi  {p_0}$, with $p_0$ the probability for the photon. Then the inequality is proved by Wootters {\it et al.} \cite {Wootters}, namely,
%%%%%%%%%%%%%%%%%%%%%%%%%%%%%%%%
\begin{align}
{{\rm P}^2} + {\Delta ^2} \le 1.
\label{22}
\end{align}
%%%%%%%%%%%%%%%%%%%%%%%%%%%%%%%%
In the case of $\Delta = 0$, a full particle behavior will appear, meaning the wave behavior disappears (${\rm P} = 1$), and vice versa.

In general, any photon will inevitably interact with the environment in the interferometer. Some information might be revealed by measuring $E$, the path distinguishability is written as
%%%%%%%%%%%%%%%%%%%%%%%%%%%%%%%%
\begin{align}
\mathcal{D} = 2{p_{{\rm{guess}}}}\left( {Z|E} \right) - 1.
\label{23}
\end{align}
%%%%%%%%%%%%%%%%%%%%%%%%%%%%%%%%
Later, Jaeger {\it et al.} \cite {Jaeger} and Englert \cite {Englert1} presented an optimal version of Eq. (\ref{22})
%%%%%%%%%%%%%%%%%%%%%%%%%%%%%%%%
\begin{align}
{\mathcal{D}^2} + {\Delta ^2} \le 1.
\label{23}
\end{align}
%%%%%%%%%%%%%%%%%%%%%%%%%%%%%%%%
Wave-particle duality relations in Eqs. (\ref{22}) and (\ref{23}) are often regarded as being different from the uncertainty relation on the concept, although there are many debate. For example, D$\ddot{{\rm u}}$rr {\it et al.} \cite {D®πrr} and Busch {\it et al.} \cite {Busch} have proven connections between certain wave-particle duality relations and Robertson's  relation building on the canonical deviation. Then, Coles and his cooperators \cite {Coles} presented Eqs. (\ref{22}) and (\ref{23}), and some other wave-particle duality relations are practically disguised uncertainty relations. Specifically, they correspond to the uncertainty relations
 with regard to the min-entropy and max-entropy for complementary observables. Hence, in contrast to the uncertainty relation, Eq. (\ref{22}) can be rewritten as
%%%%%%%%%%%%%%%%%%%%%%%%%%%%%%%%
\begin{align}
{S_{\min }}\left( Z \right) + \mathop {\min }\limits_{O \in  ( {X,Y} )} {S_{\max }}\left( O \right) \ge 1,
\label{24}
\end{align}
%%%%%%%%%%%%%%%%%%%%%%%%%%%%%%%%
where $\mathop {\min }_{O \in  ( {X,Y}  )} $ is minimized over all observables in the {\it x-y} plane of the Bloch sphere. Similar to Eq. (\ref{23}), it can be rewritten as
%%%%%%%%%%%%%%%%%%%%%%%%%%%%%%%%
\begin{align}
{S_{\min }}\left( {Z{\rm{|}}E} \right) + \mathop {\min }\limits_{O \in \left( {X,Y} \right)} {S_{\max }}\left( O \right) \ge 1.
\label{25}
\end{align}
%%%%%%%%%%%%%%%%%%%%%%%%%%%%%%%%
By parity of reasoning, other entropies can be employed to take  place of the min-entropy and the max-entropy. We cannot attain a rigid equivalence to the
wave-particle duality relations, but the conceptual meaning probably is similar. Taking examples,
Bosyk {\it et al.} \cite {Bosyk} employed the R$\acute{{\rm e}}$nyi entropy to formulate a wave-particle duality relation. Vaccaro \cite {Vaccaro}
took this approach using uncertainty relation based on Shannon entropy. Furthermore,
they provided a significant perception that wave and particle behaviors are associated with symmetry and asymmetry, respectively. In addition, Englert {\it et al.} \cite {Englert2} have
also explored entropic measurements for wave and particle behaviors of interferometers with beyond two paths.

\subsection{quantum  metrology}
In 2011, Giovannetti {\it et al.} \cite {Giovannetti} have showed that quantum metrology is used to
offer physical restrictions on the measurement's precision. The uncertainty relations play a significant role in building up these physical limits. In general, estimating an optical phase is interesting in quantum metrology, such as the phase shift in an interferometer. Therefore, the uncertainty relation involving the phase can be applied to quantum metrology. Although quantum metrology is deemed as a wide field \cite {Giovannetti}, here we only show some works that are related to the EUR. The Heisenberg's limit is termed as
a celebrated limit in quantum metrology showing the uncertainty in the phase estimation scales as $1/\left\langle N \right\rangle $, with denoting $\left\langle N \right\rangle $ as the mean photon number
when probing the phase. Further, Hall and his cooperators \cite {Hall1} have claimed that the Heisenberg's limit is considerably enlightening and  proved the following bound to put it on rigorous footing
%%%%%%%%%%%%%%%%%%%%%%%%%%%%%%%%
\begin{align}
\delta {\rm \Phi '} \ge k/\left\langle {N + 1} \right\rangle,
\label{26}
\end{align}
%%%%%%%%%%%%%%%%%%%%%%%%%%%%%%%%
where $k =: \sqrt {2\pi /{e^2}} $, and $ \delta \Psi '$ is dented by the root-mean-square deviation of the phase estimate by means of the actual phase $\Psi '$.
To prove Eq. (\ref{26}), we can denote the random variable $\Lambda {\rm{ := }}{\rm \Psi '}  - \Psi $ and combining it with the entropic uncertainty relation $H\left( N \right) + h\left( \Psi  \right) \ge \log_2 (2\pi) $, one can obtain
%%%%%%%%%%%%%%%%%%%%%%%%%%%%%%%%
\begin{align}
H\left( N \right) + h\left( \Lambda  \right) \ge \log_2 (2\pi).
\label{27}
\end{align}
%%%%%%%%%%%%%%%%%%%%%%%%%%%%%%%%
Then they can be proved by combining these equations with some identities that relate $h\left( \Lambda  \right)$ to $\delta {\rm \Psi '}  $ and $H\left( N \right)$ to $\left\langle {N + 1} \right\rangle $.
In 2012, Hall {\it et al.} \cite {Hall2} had investigated a general situation in which someone obtains  some prior information regarding the phase, and
derived a strict description about the Heisenberg's limit by means of the EUR.

\subsection{quantum teleportation}
The quantum-memory-assisted EUR in Eq. \eqref{eq3-3} is  also able to use to identify channel states useful for nonclassical teleportation
\cite{pra032338}. Based on the facts that the average teleportation
fidelity
%%%%%%%%%%%%%%%%%%%%%%%%%%%
\begin{equation}\label{eq4g-1}
 F_{\rm av}=\frac{1}{2}+\frac{1}{6}{\rm Tr}\sqrt{T^{\dag} T},
\end{equation}
%%%%%%%%%%%%%%%%%%%%%%%%%%%
is local unitary invariant \cite{Horodecki}, while any two-qubit state
is local unitary equivalent to Eq. \eqref{Eq. rhoab}, it was shown
geometrically that any $\rho_{AB}$ that corresponds to an improvement
of Berta \emph{et al.}'s uncertainty bound compared with that without
a quantum memory is useful for nonclassical teleportation \cite{pra032338}.
That is, whenever one observes a negative conditional entropy $S(A|B)$,
we are sure that $F_{\rm av}$ can exceed the classical limit $2/3$.

{\subsection{Quantum key distribution}
Technically speaking, a key distribution protocol is that two reliable players agree with sharing a key through communication over
 a public channel, and the key is secret and the adversary cannot eavesdrop from any channel.
 In tradition, the two honest players named as Alice and Bob try to share a key and the eavesdropper
 is named as Eve. Bennett {\it et al.}\cite{eb} first proposed quantum key distribution (QKD).
 Due to non-copy and non-cloning feature of quantum information \cite{www}, the application of
 symmetry argument is invalid when Alice and Bob share a key and communicate by a quantum
 channel. Simply speaking, no matter when the eavesdropper interacts with the channel and
 performs a measurement on a particle, her behaviors will inevitably lead to  noise during
 the  quantum communication. Thereby, they are capable of detecting and immediately aborting the protocol.

 Cerf {\it et al.} \cite{cer} and Grosshans {\it et al.}\cite{gro} first applied EURs on QKD.
 Especially, Koashi \cite{koa} established security criterion based on utilizing Maassen and Uffink's uncertainty relation shown as Eq. (\ref{Eq.2-6}).
 While EURs with a quantum memory offer a straightforward approach to formatting security criterions for quantum key distribution.

During the preparation step, it notes that the eavesdropper may interfere, thus we cannot
 know whether the two parties really partake a maximally entangled state or not when completing the preparation procedure.
Here we may suppose that an arbitrary state
${\rho _{ABE}}$ is shared by
%However, without loss of generality, after the preparation step it may be assumed that
Alice, Bob, and Eve.
$\vartheta $ denote a binary register in a fully mixed state which determines qubits to be measured in the basis $\mathcal {Q}$ or $\mathcal {R}$, and $T$
 is the output of Alice's measurement. Then we have
$S\left( {T|B\vartheta } \right) = \left( {1/2} \right)S\left( {Q|B} \right) + \left( {1/2} \right)S\left( {R|B} \right)$ and  analogously
$S\left( {T|E\vartheta } \right) = \left( {1/2} \right)S\left( {Q|E} \right) + \left( {1/2} \right)S\left( {R|E} \right)$.
Hence, the tripartite EUR with a quantum memory can be rewritten as
\begin{align}
S\left( {T|B\vartheta } \right) + S\left( {T|E\vartheta } \right) \ge {q_{MU}} = 1,
\label{37}
\end{align}
where $q_{MU}=1$ under the basis $\mathcal {Q}=\{|0\rangle\langle0|,|1\rangle\langle1|\}$ and $\mathcal {R}=\left\{|+\rangle\langle+|,|-\rangle\langle-|\right\}$.  The entropies
mentioned above can be
used to evaluate on the state ${\rho _{T\vartheta BE}}$ after performing the measurement on $A$. By performing measurement on $B$,
this will give rise to an estimate $\hat{T}$  of $T$. According to data-processing inequality \cite{fl}, it can be obtained
$S\left( {T|B\vartheta } \right) \le S\left( {T|\hat T} \right)$, thus we have
$S\left( {T|E\vartheta } \right) \ge 1 - S\left( {T|\hat T} \right)$. Thereby, it is clear that Eve's uncertainty of the $A$'s measuring result is large when
$S\left( {T|\hat T} \right)$ is small, which can describe quantification of the security criterion.

\subsection{Other applications}
Recently, there exists much effort to investigate the boundary of classical and quantum in physics and information processing, and these give rise to
quantitative measurements of quantumness, such as coherence and discord. We here divide the discussions into four parts. In the former two parts,
the quantum resources are discussed. Additionally, information locking and quantum coding are introduced in the last two sections, respectively.

\subsubsection{Quantum coherence}
In 2014, Baumgratz {\it et al.} \cite {coher} set up a configuration for measuring coherence.
By the amount of distillable maximally coherent states, there yields a special coherence measure,
the relative entropy of coherence, which can be written as \cite{cof2}
%%%%%%%%%%%%%%%%%%%%%%%%%%%%%%%%
\begin{align}
\Phi \left( {\mathcal{Z},\rho } \right) = D\left( {\rho \parallel \sum\limits_z {\left| {{\mathcal{Z}^z}} \right\rangle \left\langle {{\mathcal{Z}^z}} \right|\rho \left| {{\mathcal{Z}^z}} \right\rangle \left\langle {{\mathcal{Z}^z}} \right|} } \right).
\label{28}
\end{align}
%%%%%%%%%%%%%%%%%%%%%%%%%%%%%%%%

Coles {\it et al.} \cite {ier3} presented the connection between coherence and entropic uncertainty. Performing a projective measurement $Z$ on arbitrary systematic state, it can be given by
%%%%%%%%%%%%%%%%%%%%%%%%%%%%%%%%
\begin{align}
\Phi \left( {\mathcal{Z},\rho } \right) = S\left( {Z|E} \right),
\label{29}
\end{align}
%%%%%%%%%%%%%%%%%%%%%%%%%%%%%%%%
with a purifying system $E$. It means that the relative entropy of coherence as to a projective measurement is
equal to the measurement's uncertainty for the purifying system. $S\left( {Z|E} \right)$ is the uncertainty with
quantum memory. Thereby, quantum-memory-assisted EURs can be reinterpreted as lower bounds on the coherence
of the systemic state with respect to different measurements. Korzekwa {\it et al.} \cite {K.Korzekwa} discussed such an idea,
although they pay special attention to the perspective of dividing the total uncertainty into classical and quantum
parts presented by Luo  \cite {S.Luo}. Specifically, for a quantum state and an arbitrary projective measurement
$\mathcal{Z}{\rm{ = }} \{ {\left| {{\mathcal{Z}^z}} \right\rangle \left\langle {{\mathcal{Z}^z}} \right|}  \}$ ,
the classical uncertainty and the quantum uncertainty are given as the entropy of the state and as the
relative entropy of coherence, respectively.
%%%%%%%%%%%%%%%%%%%%%%%%%%%%%%%%
%\begin{align}
%\Phi \left( {\mathcal{Z},\rho } \right): = D\left( {\rho \parallel \sum\limits_z {\left| {{\mathcal{Z}^z}} \right\rangle \left\langle {{\mathcal{Z}^z}} \right|\rho \left| {{\mathcal{Z}^z}} \right\rangle \left\langle {{\mathcal{Z}^z}} \right|} } \right).
%\label{30}
%\end{align}
%%%%%%%%%%%%%%%%%%%%%%%%%%%%%%%%
Total uncertainty can be written as the summation of the classical and quantum uncertainties
%%%%%%%%%%%%%%%%%%%%%%%%%%%%%%%%
\begin{align}
S\left( Z \right) = Q\left( {\mathcal{Z},\rho } \right) + C\left( {\mathcal{Z},\rho } \right).
\label{31}
\end{align}
%%%%%%%%%%%%%%%%%%%%%%%%%%%%%%%%
Up to now, there are several uncertainty relations
related to the derived quantum uncertainty. According to Eq. (\ref {29}), these relations as EURs can be reinterpreted in the presence of a quantum memory.

\subsubsection{Information locking}
DiVincenzo {\it et al.} \cite {DiVincenzo} introduced an operational way of insight into the EUR based on
information locking. Fawzi {\it et al.} \cite {Fawzi} had discussed a cryptographic viewpoint on information locking. Basically, a locking protocol is regarded as that encoding the required classical information
onto a quantum state by means of a classical key whose magnitude is smaller than the information. In addition, it is possible to unlock and completely recover the information when the key is known.
Through  the Maassen-Uffink bound with regard to the $n$-qubit BB84 measurements, we can show the connection between information locking and entropic uncertainty
%%%%%%%%%%%%%%%%%%%%%%%%%%%%%%%%
\begin{align}
S\left( {{K^n}|{\Theta ^n}} \right) \ge n \cdot \frac{1}{2},
\label{35}
\end{align}
%%%%%%%%%%%%%%%%%%%%%%%%%%%%%%%%
where ${\Theta ^n} \in \left\{ {{\theta _1}...,{\theta _{{2^n}}}} \right\}$. One can choose randomly a $n-$qubit BB84 basis ${\theta _i}$ (key)
and encode the classical information into that basis to encode a uniformly random $n$-bit string $X$. Based on Eq. (\ref{35}), it is discussed that the mutual information
between the outcome of the measurement and the information $X$ is at the utmost $n/2$ for any measurement on quantum state \cite {DiVincenzo}. That is to say, without the knowledge of the basis choice, $n/2$ classical bits
are locked into the quantum state and be inaccessible. With this feature in mind, this is fabulous on account that
a $n$-bit or more string message is indispensable in course of
any nontrivial purely classical encryption. The question regarding the optimized trade-off between the lockable bits' amount and the key's magnitude is then raised. To do so, Fawzi {\it et al.} \cite {Fawzi} utilized the uncertainty relation
%%%%%%%%%%%%%%%%%%%%%%%%%%%%%%%%
\begin{align}
S\left( {K|\Theta } \right) \ge n \cdot \left( {1 - 2\varepsilon } \right) - {H_\mathrm{bin}}\left( \varepsilon  \right),
\label{36}
\end{align}
%%%%%%%%%%%%%%%%%%%%%%%%%%%%%%%%
where $\Theta  \in \left\{ {{\theta _1}...,{\theta _L}} \right\}$. Eq. (\ref{36}) is the basis for the information locking schemes. By virtue of the alleged canonical uncertainty relations,
state-of-the-art approach is to employ stronger definitions on information locking by the trace norm rather than mutual information \cite {Dupuis}. In final,  it is noted that Guha {\it et al.} \cite {Guha} also investigated the information locking capacity for a quantum channel, which is closely relevant with the uncertainty relation.

\subsubsection{Quantum coding}
Motivated by applications in quantum Shannon theory, some EURs with a
quantum memory were originally proposed and verified \cite {Renes1, Renes2}. Recently, Renes and his coworkers \cite {Renes3} have used the EURs as well as the equality conditions to explore the performance of quantum polar codes.

\section{Conclusions} %\label{sec:6}

Staring from Heisenberg uncertainty principle, we have reviewed the history of the entropic uncertainty relations and the recent progresses for entropy-based uncertainty relations,
with and without a quantum memory. We first recalled the development of the uncertainty relation without any quantum memory from the prospective of variance, entropy, and Majorization technology, respectively. For the methodology based on variance, Maccone and Pati \cite{Lorenzo} made a significant improvement to the conventional standard deviation which erases the shortcoming arising from the state dependency of the lower bound. Second, we reviewed the improved lower bounds for the EUR in the presence of a quantum memory. Particularly, the generalized EURs and coherence uncertainty relations were discussed. Furthermore, we provided the recent explorations on the dynamics of the entropic uncertainty within different frameworks, including open systems, curved space-time, and various noisy environments. By linking the characters of the systems to the magnitude of the uncertainty, the dynamics of the uncertainty are dynamically featured by  shape and characteristic of a system
to be observed. Lastly, we supplied the exploited applications of the EURs on various quantum tasks, which definitely give rise to the nontrivial promotion on the current and prospective quantum technologies.

To our knowledge,   there still are several challenging and open questions remaining unresolved, lying in that (i) whether there are alternative working approaches to scale the uncertainty in addition to the traditional standard deviation,
entropy and Majorization technology; (ii) As for the entropy approach of uncertainty measure, what the optimal lower bound will be, which is beyond the existed ones; (iii)
Whether the entropic uncertainty relations can be applied to more and broader topics in the field of quantum information processing;
(iv) Are there any inherent connections between EURs and Bell inequality for identifying the boundary of quantum and classical limits? If yes, how the relationship between them is like.
Thereby, with regard to EURs in the presence of quantum memory,
we look forward to receiving more attention and gaining some new and insightful results related to this theme in the future.

\begin{acknowledgements}
Wang and Ye were supported by the NSFC under Grant Nos. 61601002 and 11575001, the Fund from CAS Key Laboratory of Quantum Information under Grant No. KQI201701, and Anhui Provincial Natural Science Foundation under Grant No. 1508085QF139. Hu was supported by the NSFC under Grant No. 11675129, the New Star Project of Science and Technology of Shaanxi Province under Grant No. 2016KJXX-27, and the fund from New Star Team of XUPT.
\end{acknowledgements}

\newcommand{\PRL}{\emph{Phys. Rev. Lett.} }
\newcommand{\RMP}{\emph{Rev. Mod. Phys.} }
\newcommand{\PRA}{\emph{Phys. Rev. A} }
\newcommand{\PRB}{\emph{Phys. Rev. B} }
\newcommand{\PRE}{\emph{Phys. Rev. E} }
\newcommand{\APL}{\emph{Appl. Phys. Lett.} }
\newcommand{\NJP}{\emph{New J. Phys.} }
\newcommand{\JPA}{\emph{J. Phys. A} }
\newcommand{\JPB}{\emph{J. Phys. B} }
\newcommand{\OC}{\emph{Opt. Commun.} }
\newcommand{\PLA}{\emph{Phys. Lett. A} }
\newcommand{\EPJD}{\emph{Eur. Phys. J. D} }
\newcommand{\NP}{\emph{Nat. Phys.} }
\newcommand{\NC}{\emph{Nat. Commun.} }
\newcommand{\EPL}{\emph{Europhys. Lett.} }
\newcommand{\AoP}{\emph{Ann. Phys.} }
\newcommand{\QIC}{\emph{Quantum Inf. Comput.} }
\newcommand{\QIP}{\emph{Quantum Inf. Process.} }
\newcommand{\CPB}{\emph{Chin. Phys. B} }
\newcommand{\IJTP}{\emph{Int. J. Theor. Phys.} }
\newcommand{\IJMPB}{\emph{Int. J. Mod. Phys. B} }
\newcommand{\PR}{\emph{Phys. Rep.} }
\newcommand{\SR}{\emph{Sci. Rep.} }
\newcommand{\LPL}{\emph{Laser Phys. Lett.} }
\newcommand{\OEE}{\emph{Opt. Exp.} }
\newcommand{\IJQI}{\emph{Int. J. Quantum Inf.} }
\newcommand{\PS}{\emph{Phys. Scr.} }
\newcommand{\ADP}{\emph{Ann. Phys. (Berlin)} }

%%% Use the following two code lines if you wish to generate your bibliography with BibTeX;
%%% please replace first the string "demo" below with the name(s) of
%%% the BibTeX data base(s) you want to use.
%%% The resulting bibliography-output (the contents of the .bbl file)
%%% must be pasted into this file before submission.
%%%
%%% \bibliographystyle{andp2012}
%%% \bibliography{demo}

\begin{thebibliography}{000}
%%% The number of zeroes here should correspond to the number of digits in the
%%% number of bibliography entries here: if there are up to 9 entries, put one
%%% zero; if there 10 up to 99 entries, put two zeroes; and so on.







%% 1-10
\bibitem {Heisenberg} W. Heisenberg, \emph{Z. Phys. } \textbf{1927}, \emph{43}, 172.
\bibitem {E.H.Kennard} E. H. Kennard, \emph{Z. Phys.} \textbf {1927}, \emph{44}, 326.
\bibitem {PatrickJ} P. J. Coles, M. Berta, M. Tomamichel,  S. Wehner, \emph{Rev. Mod. Phys.} \textbf{2017}, \emph{87}, 015002.
\bibitem {Everett1} H. Everett, \emph{Rev. Mod. Phys.} {\bf 1957}, \emph{29}, 454.
\bibitem {Hirschman} I. I. Hirschman, \emph{Am. J. Math.} {\bf1957}, \emph{79}, 152.
\bibitem {Beckner}  W. Beckner, \emph{Ann. Math.} {\bf 1975}, \emph{102}, 159.
\bibitem {Birula} I. Bia${\l}$ynicki-Birula, J. Mycielski,  \emph{Commun. Math. Phys.} {\bf 1975}, \emph{44}, 129.
\bibitem {Birula2} I. Bia${\l}$ynicki-Birula, ${\l}$. Rudnicki, {\bf 2011}, in Statistical Complexity, edited by K. Sen (Springer Netherlands, Dordrecht).
\bibitem {Beckner2} S. Wehner,  A. Winter, \NJP {\bf 2010}, \emph{12}, 025009.
{\bibitem {hc1} A. Hertz, N. J Cerf,  \emph{J. Phys. A: Math. Theor.} {\bf 2019}, \emph{52}, 173001.}
\bibitem {H.P.Robertson} H. P. Robertson, \emph{Phys. Rev. } \textbf{1929}, \emph{34}, 163.

%%% 11-20
\bibitem{qiao} J. L. Li, C. F. Qiao, \emph{J. Phys. A: Math. Theor.} {\bf 2017}, \emph{50}, 03LT01.
\bibitem {Schr}  E. Schr\"{o}dinger,  \emph{Physikalisch-mathematische Klasse} \textbf{1930}, \emph{14}, 296.
\bibitem{Lorenzo} L. Maccone, A. K. Pati, \PRL \textbf{2014}, \emph{113}, 260401.
\bibitem{Kunkun} K. K. Wang, X. Zhan, Z. H. Bian, J. Li, Y. S. Zhang, P. Xue, \emph{Phys. Rev. A} \textbf{2016}, \emph{93}, 052108.
\bibitem{Kunkun2} L. Xiao, K. Wang, X. Zhan, Z. Bian, J. Li, Y. Zhang, P. Xue, A. K. Pati, \OEE {\bf 2017}, \emph{25}, 17904.
\bibitem{Kunkun3} B. Fan, K. K. Wang, L. Xiao,  P. Xue, \PRA \textbf{2018}, \emph{98}, 032118.
\bibitem {D.Deutsch} D. Deutsch, \emph{Phys. Rev. Lett.} {\bf 1983}, \emph{50}, 631.
\bibitem {K.Kraus} K. Kraus, \emph{Phys. Rev. D} {\bf 1987}, \emph{35}, 3070.
\bibitem{H.Maassen} H. Maassen, J. B. M. Uffink, \emph{Phys. Rev. Lett. } {\bf 1988}, \emph{60}, 1103.
\bibitem{K.Korzekwa} K. Korzekwa, M. Lostaglio, D. Jennings, T. Rudolph, \PRA \textbf{2014}, \emph{89}, 042122.
\bibitem{RE} A. R\'{e}nyi, in Proceedings of the 4th Berkeley Symposium on Mathematical Statistics and Probability, Vol. 1 (University of California Press, Berkeley, CA), \textbf{1960}, pp. 547-561.
{ \bibitem{bk1}K. Baek, H. Nha, W. Son, \emph{Entropy} {\bf2019}, \emph{21}, 270.}

\bibitem{Ghasemi} A. Ghasemi, M. R. Hooshmandasl, M. K. Tavassoly, \emph{Phys. Scr.} {\bf2011}, \emph{84}, 035007.
%%% 21-30
\bibitem{Dodonov1} V. V. Dodonov, A. V. Dodonov, \emph{Phys. Scr.} {\bf 2015}, \emph{90}, 074049.
\bibitem{Rastegin} A. E. Rastegin, \ADP {\bf 2019}, \emph{531}, 1800466.
\bibitem{Pegg1}D. T. Pegg, \PRA {\bf 1998}, \emph{58}, 4307.
\bibitem{Partovi} M. H. Partovi, \emph{Phys. Rev. A} \textbf{2011}, \emph{84}, 052117.
\bibitem{Friedland} S. Friedland, V. Gheorghiu, G. Gour, \emph{Phys. Rev. Lett.} \textbf{2013}, \emph{111}, 230401.
\bibitem{Pucha} Z. Pucha{\l}a, {\L}. Rudnicki, K. \.{Z}yczkowski, \emph{J. Phys. A} \textbf{2013}, \emph{46}, 272002.
\bibitem{Renes} J. M. Renes, J. C. Boileau, \PRL \textbf{2009}, \emph{103}, 020402.
\bibitem{Berta} M. Berta, M. Christandl, R. Colbeck, J. M. Renes, R. Renner, \NP \textbf{2010}, \emph{6}, 659.
\bibitem{Nphys1} C. F. Li, J. S. Xu, X. Y. Xu, K. Li, G. C. Guo, \NP \textbf{2011}, \emph{7}, 752.
\bibitem{Nphys2} R. Prevedel, D. R. Hamel, R. Colbeck, K. Fisher, K. J. Resch, \NP \textbf{2011}, \emph{7}, 757.

%%% 31-40
\bibitem{APL} Z. Y. Xu, S. Q. Zhu, W. L. Yang, \APL \textbf{2012}, \emph{101}, 244105.
\bibitem{Pati} A. K. Pati, M. M. Wilde, A. R. Usha Devi, A. K. Rajagopal, Sudha, \PRA \textbf{2012}, \emph{86}, 042105.
\bibitem{QD} H. Ollivier, W. H. Zurek, \PRL \textbf{2001}, \emph{88}, 017901.
\bibitem{pra014105} M. L. Hu, H. Fan, \PRA \textbf{2013}, \emph{88}, 014105.
\bibitem{score} M. N. Bera, R. Prabhu, A. Sen(De), U. Sen, \PRA \textbf{2012}, \emph{86}, 012319.
\bibitem{bound1} P. J. Coles, M. Piani, \PRA \textbf{2014}, \emph{89}, 022112.
\bibitem{bound2} F. Adabi, S. Salimi,  S. Haseli, \PRA \textbf{2016}, \emph{93}, 062123.
 {\bibitem{hs22}S. Haseli, F. Ahmadi, \emph{Eur. Phys. J. D}  \textbf{2019}, \emph{73}, 65.}
\bibitem{LMF} S. Liu, L. Z. Mu, H. Fan, \PRA \textbf{2015}, \emph{91}, 042133.
\bibitem{Yucs} J. Zhang, Y. Zhang, C. S. Yu, \SR \textbf{2015}, \emph{5}, 11701.
\bibitem{qip1} H. Dolatkhah, S. Haseli, S. Salimi, A. S. Khorashad, \QIP \textbf{2019}, \emph{18}, 13.

%%% 41-50
\bibitem{pra022314} M. L. Hu, H. Fan, \PRA \textbf{2013}, \emph{87}, 022314.
\bibitem{Nielsen} M. A. Nielsen, I. L. Chuang, \emph{Quantum Computation and Quantum Information (Cambridge University Press, Cambridge, 2000)}.
\bibitem{EoF1} C. H. Bennett, D. P. DiVincenzo, J. A. Smolin, W. K. Wootters, \PRA \textbf{1996}, \emph{54}, 3824.
\bibitem{EoF2} W. K. Wootters, \PRL \textbf{1998}, \emph{80}, 2245.
\bibitem{Koashi} M. Koashi, A. Winter, \PRA \textbf{2004}, \emph{69}, 022309.
\bibitem{ulqe} F. Buscemi, G. Gour, J. S. Kim, \PRA \textbf{2009}, \emph{80}, 012324.
\bibitem{ulqd} Z. Xi, H. Fan, Y. Li, \PRA \textbf{2012}, \emph{85}, 052102.
\bibitem{PhysRep} M. L. Hu, X. Hu, J. Wang, Y. Peng, Y. R. Zhang, H. Fan, \PR \textbf{2018}, \emph{762}, 1.
\bibitem{Hupra} M. L. Hu, H. Fan, \PRA  \textbf{2017}, \emph{95}, 052106.
\bibitem{eurqc2} X. Yuan, G. Bai, T. Peng, X. Ma, \PRA \textbf{2017}, \emph{96}, 032313.
\bibitem{coher} T. Baumgratz, M. Cramer, M. B. Plenio, \PRL \textbf{2014}, \emph{113}, 140401.
{ \bibitem{Malvezzi22}  A. L. Malvezzi, 1 G. Karpat,B. \c{C}akmak,  F. F. Fanchini, T. Debarba,  R. O. Vianna, \PRB \textbf{2016}, \emph{93}, 184428.}
{ \bibitem{he22} J. He, Z. Y. Ding, J. D. Shi,   T. Wu, \ADP \textbf{2018}, \emph{530}, 1800167.}
{ \bibitem{HML11} M. L. Hu, W, Zhou, \emph{Laser Phys. Lett. }  {\bf2019}, \emph{16}, 045201.}
%%% 51-60
\bibitem{cof1} X. Yuan, H. Zhou, Z. Cao, X. Ma, \PRA  \textbf{2015}, \emph{92}, 022124.
\bibitem{cof2} A. Winter, D. Yang, \PRL \textbf{2016}, \emph{116}, 120404.
\bibitem{eurqc3} U. Singh, A. K. Pati, M. N. Bera, \emph{Mathematics} \textbf{2016}, \emph{4}, 47.
\bibitem{ier1} M. J. W. Hall, \PRL \textbf{1995}, \emph{74}, 3307.
\bibitem{ier2} M. J. W. Hall, \PRA \textbf{1997}, \emph{55}, 100.
\bibitem{ier3} P. J. Coles, L. Yu, V. Gheorghiu, R. B. Griffiths, \PRA \textbf{2011}, \emph{83}, 062338.
\bibitem{ier4} A. Grudka, M. Horodecki, P. Horodecki, R. Horodecki, W. K{\l}obus, {\L}. Pankowski, \PRA \textbf{2013}, \emph{88}, 032106.
\bibitem {Z.Y.Xu} Z. Y. Xu, W. L. Yang, M. Feng, \emph{Phys. Rev. A } {\bf 2012}, \emph{86}, 012113.
%Quantum-memory-assisted entropic uncertainty relation under noise.
\bibitem {D.Wang} D. Wang, W. N. Shi, R. D. Hoehn, F. Ming, W. Y. Sun, S. Kais, L. Ye, \emph{Ann. Phys. (Berlin)} {\bf 2018}, \emph{530}, 1800080.
%Effects of Hawking radiation on the entropic uncertainty in a Schwarzschild space-time.

%%% 61-70
\bibitem {A.J.Huang} A. J. Huang, J. D. Shi, D. Wang, L. Ye, \emph{Quantum Inf. Process.} {\bf 2017}, \emph{16}, 46.
%Steering quantum-memory-assisted entropic uncertainty under unital and nonunital noises via filtering operations.
\bibitem {Ming1} F. Ming, D. Wang, A. J. Huang, W. Y. Sun, J. D. Shi, L. Ye, \emph{Quantum Inf. Process.} {\bf 2018}, \emph{17}, 9.
{\bibitem {kar2}G. Karpat, \emph{Can. J. Phys.} {\bf2018}, \emph{96}, 700.}
{\bibitem {wgy1}G. Y. Wang, Y. N. Guo,  K. Zeng, \emph{J. Mod. Opt.} {\bf2019}, \emph{66}, 367.}
{\bibitem {jyh1} Y. H. Ji, Q. Ke,  J. J. Hu, \emph{Physica  E}  {\bf2019}, \emph{110}, 140.}

\bibitem {D.Wang1} D. Wang, F. Ming, A. J. Huang, W. Y. Sun, J. D. Shi, L. Ye, \emph{Sci. Rep.} {\bf 2017}, \emph{7}, 1066.
%Entropic uncertainty relations for Markovian and non-Markovian processes under a structured bosonic reservoir.
\bibitem {D.Wang2} D. Wang, W. N. Shi, F. Ming, R. D. Hoehn, W. Y. Sun, L. Ye, S. Kais, \emph{Quantum Inf. Process.} {\bf 2018}, \emph{17}, 335.
%Probing entropic uncertainty relations under a two-atom system coupled with structured bosonic reservoirs.
\bibitem {S.Maniscalco}	S. Maniscalco, F. Petruccione, \emph{Phys. Rev. A} {\bf 2006}, \emph{73}, 012111.
%Non-Markovian dynamics of a qubit.
\bibitem{Huaop2343} M. L. Hu, H. Fan, \AoP \textbf{2012}, \emph{327}, 2343.
\bibitem {M.N.Chen1} M. N. Chen, D. Wang, L. Ye, \emph{Phys. Lett. A} \textbf{2019}, \emph{383}, 977.
{\bibitem {kar1}G. Karpat, J. Piilo, S. Maniscalco, \emph{EPL} {\bf2015}, \emph{111}, 50006.}
{\bibitem {cpf222}P. F. Chen, L. Ye, D. Wang, \emph{Eur. Phys. J. D} {\bf2019}, DOI: 10.1140/epjd/e2019-100013-0.}

\bibitem {J.Feng1} J. Feng, Y. Z. Zhang, M. D. Gould, H. Fan, \emph{Phys. Lett. B} {\bf 2015}, \emph{743}, 198.
%Uncertainty relation in Schwarzschild spacetime.
\bibitem {J.L.Huang} J. L. Huang, F. W. Shu, Y. L. Xiao, M. H. Yung, \emph{Eur. Phys. J. C} {\bf 2018}, \emph{78}, 545.
%Holevo bound of entropic uncertainty in Schwarzschild spacetime.
\bibitem {Z.Y.Zhang1} Z. Y. Zhang, J. M. Liu, Z. F. Hu, Y. Z. Wang, \emph{Ann. Phys. (Berlin}) {\bf 2018}, \emph{530}, 1800208.

{\bibitem {ming22}F. Ming, D. Wang, L. Ye, \emph{ Ann. Phys. (Berlin)} {\bf2019}, DOI: 10.1002/andp.201900014.}


\bibitem {vg} G. Vidal, R. F. Werner, \emph{Phys. Rev. A} {\bf 2002}, \emph{65}, 032314.
%%% 71-80
\bibitem  {D.Wang3} D. Wang, F. Ming, A. J. Huang, W. Y. Sun, J. D. Shi, L. Ye, \emph{Laser Phys. Lett.} {\bf 2017}, \emph{14}, 055205.
%Exploration of quantum-memory-assisted entropic uncertainty relations in a noninertial frame.
\bibitem {A.J.Huang3}A. J. Huang, D. Wang, J. M. Wang, J. D. Shi, W. Y. Sun, L. Ye, \emph{Quantum Inf. Process.} {\bf 2017}, \emph{16}, 204 .
% Exploring entropic uncertainty relation in the Heisenberg XX model with inhomogeneous magnetic ?eld.
\bibitem {D.Wang5} D. Wang, F. Ming, A. J. Huang, W. Y. Sun, L. Ye, \emph{Laser Phys. Lett.} {\bf 2017},  \emph{14}, 095204.
\bibitem {Ming2} {F. Ming, D. Wang, W. N. Shi, A. J. Huang, W. Y. Sun, L. Ye, \emph{Quantum Inf. Process.} {\bf 2018}, \emph{17}, 89.}
\bibitem {D.Wang4} D. Wang, A. J. Huang, F. Ming, W. Y. Sun, H. P. Lu, C. C. Liu, L. Ye, \emph{Laser Phys. Lett.} {\bf 2017},  \emph{14}, 065203.
\bibitem {X.Zheng} X. Zheng, G. F. Zhang, \QIP {\bf 2017}, \emph{16}, 1.
{\bibitem {hzm11}Z. M. Huang, \emph{Laser Phys. Lett. } {\bf 2018}, \emph{15}, 025203.}
\bibitem {Ming3}  {F. Ming, D. Wang, W. N. Shi, A. J. Huang, M. M. Du, W. Y. Sun, Liu Ye, \emph{Quantum Inf. Process.} {\bf 2018}, \emph{17}, 267.}
\bibitem {Y.Y.Yang} Y. Y. Yang, W. Y. Sun, W. N. Shi, F. Ming, D. Wang, L. Ye, \emph{Front. Phys.} {\bf 2019},  \emph{14}, 31601.

{\bibitem {zhangz22} Z. Y. Zhang, D. X. Wei, J. M. Liu, \LPL  {\bf 2018}, \emph{15}, 065207.}

{\bibitem {swn11}W. N. Shi, F. Ming, D. Wang, L. Ye, \emph{Laser Phys. Lett. } {\bf 2019}, \emph{18}, 70.}
\bibitem {Y.L.Zhang} Y. L. Zhang, M. F. Fang, G. D Kang, Q. P. Zhou, \emph{Quantum Inf. Process.} {\bf 2018}, \emph{17}, 62.
{\bibitem {cpf22} P. F. Chen, W. Y. Sun, F. Ming, A. J. Huang, D. Wang, L. Ye, \emph{Laser Phys. Lett.} {\bf 2019}, \emph{15}, 015206.}
{\bibitem {hs11}S. Haseil, H Dolatkhah, S Salimi, A S Khorashad, \emph{Laser Phys. Lett.} {\bf 2019}, \emph{16}, 045207.}

\bibitem {Y.N.Guo}  Y. N. Guo, M. F. Fang, Q. L. Tian, Z. D. Li, K. Zeng, \emph{Laser Phys. Lett.} {\bf 2018}, \emph{15}, 105205.
\bibitem {Q.Su} Q. Su, M. Al-Amri, L. Davidovich, M. Suhail Zubairy, \emph{Phys. Rev. A} {\bf 2010}, \emph{82}, 052323.
% Reversing entanglement change by a weak measurement.
\bibitem {Bender} C. M. Bender, S. Boettcher, \emph{Phys. Rev. Lett.} {\bf 1988}, \emph{80}, 5243.
%Real Spectra in Non-Hermitian Hamiltonians Having $\mathcal{P T}$ Symmetry.
\bibitem {W.N.Shi}  W. N. Shi, D. Wang, W. Y. Sun, F. Ming, A. J. Huang, L. Ye, \emph{Laser Phys. Lett.} {\bf 2018}, \emph{15}, 075202.
%Steering the measured uncertainty under decoherence through local P-T symmetric operations.
\bibitem {Yu1}M. Yu, M. F. Fang, \QIP {\bf2017}, \emph{16}, 213.
{ \bibitem {adabi11}F. Adabi, S. Haseli, S. Salimi, \emph{EPL } {\bf2016}, \emph{115}, 60004.}

\bibitem {G®πhne} O. G$\ddot{{\rm u}}$hne, G. T$\acute{{\rm o}}$th, \emph{Phys. Rep.} {\bf 2009}, \emph{474}, 1.
\bibitem {Horodecki1} R. Horodecki, P. Horodecki, M. Horodecki,  K. Horodecki, \emph{Rev. Mod. Phys.} {\bf 2009}, \emph{81}, 865.
\bibitem {R.Namiki} R. Namiki, Y. Tokunaga, \emph{Phys. Rev. Lett.} {\bf 2012}, \emph{108}, 230503.
\bibitem {VG}  V. Giovannetti, \emph{Phys. Rev. A} {\bf 2004}, \emph{70}, 012102.
\bibitem {OG}  O. G\"{u}hne, M. Lewenstein, \emph{Phys. Rev. A} {\bf 2004}, \emph{70}, 022316.
\bibitem {Hu112} Huang, Y., \emph{Phys. Rev. A}   {\bf 2010}, \emph{82}, 012335.
\bibitem{Horodecki} R. Horodecki, M. Horodecki, P. Horodecki, \PLA \textbf{1996}, \emph{222}, 21.
\bibitem {M.Berta2}	M. Berta, P. J. Coles, S. Wehner, \emph{Phys. Rev. A} {\bf 2014}, \emph{90}, 062127.
\bibitem {S.P.Walborn} S. P. Walborn, B. G. Taketani, A. Salles, F. Toscano, R. L. de Matos Filho, \emph{Phys. Rev. Lett.} {\bf 2009}, \emph{103}, 160505.

%%% 91-100
\bibitem {A.Saboia} A. Saboia, F. Toscano, S. P. Walborn, \emph{Phys. Rev. A } {\bf 2011}, \emph{83}, 032307.
\bibitem {Y.Huang}  Y. Huang,  \emph{IEEE Trans. Inf. Theory} {\bf 2013}, \emph{59}, 6774.
\bibitem {Wiseman1} H. M. Wiseman,S. J. Jones, A. C. Doherty, \emph{Phys. Rev. Lett.} {\bf 2007},  \emph{98}, 140402.
\bibitem {E.G.Cavalcanti} E. G. Cavalcanti, S. J. Jones, H. M.Wiseman, M. D. Reid,  \emph{Phys. Rev. A} {\bf 2009}, \emph{80}, 032112.
\bibitem {J.Schneeloch} J. Schneeloch, C. J. Broadbent, S. P. Walborn, E. G. Cavalcanti, J. C. Howell, \emph{Phys. Rev. A} {\bf 2013}, \emph{87}, 062103.
\bibitem {Wootters} W. Wootters, W. H. Zurek, \emph{Phys. Rev. D} {\bf 1979}, \emph{19}, 473.
\bibitem {Jaeger} G. Jaeger, A. Shimony, L. Vaidman, \emph{Phys. Rev. A} {\bf 1995}, \emph{51}, 54.
\bibitem {Englert1} B. G. Englert, \emph{Phys. Rev. Lett.} {\bf 1996}, \emph{77}, 2154.
\bibitem {D®πrr} S. D$\ddot{{\rm u}}$rr, G. Rempe,  \emph{Am. J. Phys.} {\bf 2000}, \emph{68}, 1021.
\bibitem {Busch} P. Busch, C. Shilladay, \emph{Phys. Rep.} {\bf 2006}, \emph{435}, 1.

%%% 101-110
\bibitem {Coles} P. J. Coles, J. Kaniewski, S. Wehner, \emph{Nat. Commun.} {\bf 2014}, \emph{5}, 5814.
\bibitem {Bosyk} G. M. Bosyk, M. Portesi, F. Holik, A. Plastino, \emph{Phys. Scr.} {\bf 2013}, \emph{87}, 065002.
\bibitem {Vaccaro} J. A. Vaccaro, \emph{Proc. R. Soc. A} {\bf 2011}, \emph{468}, 1065.
\bibitem {Englert2} B. G. Englert, D. Kaszlikowski, L. C. Kwek, W. H. Chee, \emph{Int. J. Quantum Inform.} {\bf 2008}, \emph{06}, 129.
\bibitem {Giovannetti}  V. Giovannetti, S. Lloyd, L. Maccone, \emph{Nat. Photonics} {\bf 2011}, \emph{5}, 222.
\bibitem {Hall1} M. J. W. Hall, D. W. Berry, M. Zwierz, H. M. Wiseman, \emph{Phys. Rev. A} {\bf 2012}, \emph{85}, 041802.
\bibitem {Hall2}M. J. W. Hall, H. M. Wiseman, \NJP {\bf 2012}, \emph{14}, 033040.
\bibitem{pra032338} M. L. Hu, H. Fan, \PRA \textbf{2012}, \emph{86}, 032338.
\bibitem{eb} C. H. Bennett, G. Brassard, in Proceedings of the IEEE International Conference on Computers, Systems and Signal Processing 1984,  {\bf 1984}, Vol. 1 (IEEE, Bangalore), pp. 175-179.
\bibitem{www} W. K. Wootters, W. H. Zurek, \emph{Nature (London)} {\bf1982},  \emph{299}, 802.
\bibitem{cer} N. J. Cerf, M. Bourennane, A. Karlsson,  N. Gisin, \emph{Phys. Rev. Lett.} {\bf 2002}, \emph{88,} 127902.
\bibitem{gro} F. Grosshans,  N. J. Cerf, \emph{Phys. Rev. Lett.} {\bf 2004}, \emph{92,} 047905.
\bibitem{koa} M. Koashi, \emph{J. Phys. Conf. Ser.}  {\bf 2006}, \emph{36,} 98.
\bibitem{fl}R. L. Frank, E. H. Lieb, \emph{J. Math. Phys. (N.Y.)} {\bf 2013}, \emph{54}, 122201.



\bibitem{S.Luo} S. Luo,  \emph{Theor. Math. Phys. } \textbf{2005}, \emph{143}, \emph{681}.
\bibitem {DiVincenzo}D. DiVincenzo, M. Horodecki, D. Leung, J. Smolin, B. Terhal, \emph{Phys. Rev. Lett.} {\bf 2004}, \emph{92}, 067902.

%%% 111-116
\bibitem {Fawzi} O. Fawzi, P. Hayden, P. Sen, \emph{in Proceedings of ACM STOC 2011, (ACM Press, New York),} {\bf 2011}, pp. 773-782.
\bibitem {Dupuis} F. Dupuis, J. Florjanczyk, P. Hayden, D. Leung, \emph{Proc. R. Soc. A} {\bf 2013}, \emph{469}, 20130289.
\bibitem {Guha} S. Guha, P. Hayden, H. Krovi, S. Lloyd, C. Lupo, J. H. Shapiro, M. Takeoka, M. M. Wilde, \emph{Phys. Rev. X} {\bf 2014}, \emph{4}, 011016.
\bibitem {Renes1} J. M. Renes, J. C. Boileau, \emph{Phys. Rev. A} {\bf 2008}, \emph{78}, 032335.
\bibitem {Renes2} J. M. Renes, M. M. Wilde, \emph{IEEE Trans. Inf. Theory} {\bf 2014}, \emph{60}, 3090.
\bibitem {Renes3} J. M. Renes, D. Sutter, F. Dupuis, R. Renner, \emph{IEEE Trans. Inf. Theory} {\bf 2015}, \emph{61}, 6395.

\end{thebibliography}
%%%
%%% If you are doing it by hand make sure to put the correct make up into
%%% it.  Please see below for the usual keys; and please keep in mind that all the custom
%%% macros (\jr, \othercit} and remarkably also \textsc  here have no effect
%%% regarding typesetting.  They are crucial for the hypertext markup of the
%%% data, though.
%%%
%%% The macros are:
%%% \textsc for authors' names;  also for editors' names, if there are no authors;
%%% \jr for (abbreviated) journal names;
%%% \othercit to be used as a prefix to \bibitem in non-journal entries and like
%%% a common makro for partial entries in multi-entry \bibref-constructs.
%%%
%%% Replace the following example bibliography with your references
%%% before submission:

\end{document}